\documentclass[letterpaper]{article} 
\usepackage{aaai25}  
\usepackage{times}  
\usepackage{helvet}  
\usepackage{courier}  
\usepackage[hyphens]{url}  
\usepackage{graphicx} 
\usepackage{natbib}  
\usepackage{caption} 
\usepackage{algorithm}
\usepackage{algorithmic}

\usepackage{newfloat}
\usepackage{listings}
\DeclareCaptionStyle{ruled}{labelfont=normalfont,labelsep=colon,strut=off} 
\floatstyle{ruled}
\newfloat{listing}{tb}{lst}{}
\floatname{listing}{Listing}
\usepackage{booktabs}
\usepackage[group-separator={,}]{siunitx}
\usepackage{cleveref}

\usepackage{tabularx}

\usepackage{multirow}

\usepackage{subcaption}

\usepackage{xcolor}
\newcommand{\answerYes}[1]{\textcolor{blue}{#1}} 
\newcommand{\answerNo}[1]{\textcolor{teal}{#1}} 
\newcommand{\answerNA}[1]{\textcolor{gray}{#1}} 
 
\title{
Studying Behavioral Addiction by Combining Surveys and Digital Traces:\\A Case Study of TikTok\footnote{\textcolor{red}{Accepted at ICWSM 2025, to appear.}}
}

\author {
    Cai Yang\textsuperscript{\rm 1,2},
    Sepehr Mousavi\textsuperscript{\rm 1},
    Abhisek Dash\textsuperscript{\rm 1},
    Krishna P. Gummadi\textsuperscript{\rm 1},
    Ingmar Weber\textsuperscript{\rm 2}
}
\affiliations {
    \textsuperscript{\rm 1}Max Planck Institute for Software Systems\\
    \textsuperscript{\rm 2}I2SC, Saarland University\\
    \textsuperscript{\rm 1}\{caiyang, smousavi, adash, gummadi\}@mpi-sws.org, \textsuperscript{\rm 2}iweber@cs.uni-saarland.de
}

\begin{document}

\maketitle

\begin{abstract}

Opaque algorithms disseminate and mediate the content that users consume on online social media platforms. 
This algorithmic mediation serves users with contents of their liking, on the other hand, it may cause several inadvertent risks to society at scale. 
While some of these risks, e.g., filter bubbles or dissemination of hateful content, are well studied in the community, \textit{behavioral addiction}, designated by the Digital Services Act (DSA) as a potential systemic risk, has been understudied.
In this work, we aim to study if one can effectively diagnose behavioral addiction using digital data traces from social media platforms.
Focusing on the TikTok short-format video platform as a case study, we employ a novel mixed methodology of combining survey responses with data donations of behavioral traces.
We survey \num{1590} TikTok users and stratify them into three addiction groups (i.e., less/moderately/highly likely addicted).
Then, we obtain data donations from \num{107} surveyed participants.
By analyzing users' data we find that, among others, highly likely addicted users spend more time watching TikTok videos and keep coming back to TikTok throughout the day, indicating a compulsion to use the platform.
Finally, by using basic user engagement features, we train classifier models to identify highly likely addicted users with $F_1 \geq 0.55$. The performance of the classifier models suggests predicting addictive users solely based on their usage is rather difficult. 

\end{abstract}

\section{Introduction}
Over the past years, two major trends have transformed the social media landscape: (i) an increasing shift toward algorithm-driven personalized recommendations~\cite{narayanan2023understanding, wsj2021Investigation, Smith2021How}
; and (ii) an increase in the consumption of short-format video content~\cite{zannettou2024analyzing} on platforms like TikTok\footnote{\url{https://www.tiktok.com/}}, Instagram \footnote{\url{https://www.instagram.com/}}, YouTube Shorts \footnote{\url{https://www.youtube.com/shorts/}}, etc. 
Even if new users sign up on these platforms, within a few interactions, their deployed algorithms are able to predict and present relevant content to them, encouraging them to continue browsing.

However, the effectiveness of algorithmically curated social media platforms also comes with inadvertent systemic risks.
Some of the risks, ranging from filter bubbles~\cite{seargeant2019social} to the dissemination of extreme~\cite{ribeiro2020auditing} and hateful content~\cite{saha2023hate}, have been well studied. 
On the other hand, a critical risk that has been understudied in the literature is that of ``behavioral addiction''.

Addiction can be broadly divided into two categories: substance addiction and behavioral addiction. 
The core feature of both categories of addiction is the failure to resist an impulse, drive, or temptation to consume psychoactive substances or engage with behaviors that are harmful to the person or others~\cite{grant2010introduction}.
Behavioral addiction can be defined as a compulsion to engage with a particular behavior despite the behavior causing significant impairment or distress in several aspects of a person's life.
So far only gambling has been clinically recognized as a behavioral addiction~\cite{american2013diagnostic,holden2010behavioral}.
Concerning social media, behavioral addiction can be seen as a compulsion to engage with social media content despite its harmful effects.


It is worth emphasizing that neither substance nor behavior addiction is defined by the amount of usage, but rather by the harm induced by the compulsive substance or behavior.
For instance, alcohol addiction is \emph{not} diagnosed based on the amount of alcohol consumed, but rather by the harms such addictive consumption causes~\cite{american2013diagnostic}. 
Similarly, it is still an open question as to how much of the potential negative effects of social media addiction are due to the \emph{quality} vs. \emph{quantity} of use.

\noindent
\paragraph{Importance of studying social media addiction}
The Digital Services Act (DSA)~\cite{EC2022DSARegulation} has put forward a set of guidelines for very large online platforms (VLOPs)~\cite{EC2023VLOP} whereby they need to assess the systemic risks they may affect.
The DSA guideline itself has identified four potential systemic risks including Behavioral addiction (recital 83). 
The seriousness of policymakers toward this risk can be gauged from the recent legal proceedings that social media platforms, and in particular TikTok, are facing for addictive design practices both in the European Union~\cite{EC2024Proceedings} and in the United States of America~\cite{NYC2024Lawsuit}.

While policy makers and society, in general, expect social media platforms to take proactive measures to detect and mitigate behavioral addiction, there has been no formal study of \textit{whether 
behavioral addiction can be identified using users' digital traces} on these platforms. 
Although such digital traces are rich, harms which are essential in diagnosing behavioral addiction may or may not be included in such data.  
Hence, we intend to investigate the utility of digital traces in studying behavioral addiction from the lens of (a)~user perception and (b)~data from the platforms.
The former provides valuable insights into participants' perceptions of their social media usage and allows us to utilize prior established addiction detection methods. On the other hand, the latter enables us to evaluate the efficacy of digital traces in diagnosing behavioral addiction.

The methodology described in this paper can be adapted to any social media platform. However, due to its popularity and involvement in numerous legal proceedings related to addictive design practices, the current work focuses on TikTok.
Next, we discuss our research questions, followed by our observations and their implications.

\noindent
\textit{RQ1: Do participants suffer from behavioral addiction?}
We investigate the existence of behavioral addiction among users of TikTok through a crowd-sourced user survey. For this, we recruit \num{1590} participants on Prolific.
The participants are asked a number of questions, including a set of six questions (as per Bergen Facebook Addiction Scale~\cite{doi:10.2466/02.09.18.PR0.110.2.501-517}) to understand participants' salience, tolerance, mood modification, relapse, withdrawal, and conflict.
Additionally, we directly asked participants if they feel they are addicted to TikTok.
Based on their answers, we divide participants into three categories: (a)~highly likely, (b)~moderately likely, and (c)~less likely addicted. 

\noindent
\textit{$\bullet$ Key Observations:}
\num{436} out of \num{1590} (27\%) surveyed participants are found to be highly likely addicted.
Moreover, 39\% of participants belonging to age group $[18, 24]$ are found to be in this category. 

\noindent
\textit{RQ2: Do participants in different addiction groups exhibit different usage patterns?}
To understand the usage patterns of the participants in different addiction groups, we need their user engagement data on TikTok.
To this end, we leverage their rights to data access under GDPR, whereby users can request TikTok to provide them with their data and donate it to us (if they so choose) for further analysis.
A set of 187 respondents participated in the second part of this study.
However, after accounting for their data compliance and longitudinal checks, we study 107 data donations. 

\noindent
\textit{$\bullet$ Key Observations:}
Highly likely addicted users tend to spend more time watching videos, and have more frequent individual sessions.
In other words, highly likely addicted users keep coming back to TikTok throughout the day, showing a compulsion to use the platform.

\noindent
\textit{RQ3: Can we predict addiction level based on a user's social media data?}
Finally, to understand if one can predict addiction levels from TikTok usage data, we create feature vectors for each user and train a classifier.
The aim here is to understand how effective the features derived from usage patterns are in predicting an individual's addiction level.

\noindent
\textit{$\bullet$ Key Observations:}
Our observations indicate predicting addiction status given solely basic usage data, such as time spent on the platform, is rather difficult.
Even with more features, a multi-layer perceptron classifier performs only moderately at identifying highly likely addicted users with $F_1 \geq 0.55$.

\section{Related Work}
\label{sec:related}

\subsection{Auditing Social Media Platforms}
\label{subsec:related-audit}
Given the scale at which online social media touch the fabrics of modern society, it is imperative to understand the inadvertent consequences that may result from large-scale engagement in this algorithm-mediated environment.
Hence, auditing social media platforms has emerged as a new paradigm of research.
Multiple prior work have tried to audit and study biases on online social media~\cite{ali2019discrimination, kulshrestha2017quantifying}, algorithmic transparency~\cite{vombatkere2024tiktok}, algorithmic explanations~\cite{mousavi2024auditing}, spread of misinformation and hate speech~\cite{ribeiro2020auditing, saha2021short} to name a few motivating use-cases. 

However, there has been no prior work extensively investigating behavioral addiction in online social media, especially by analyzing user engagement data at scale.
The inclusion of behavioral addiction as part of the systemic risks of very large online platforms under DSA~\cite{EC2022DSARegulation} further underlines the timeliness of the current study to bridge the mentioned research gap.

\subsection{ Combining Surveys and Log Data} 
\label{subsec:related-survey-log}

Modeling users' behaviors solely by their self-reports raises validity concerns.
For instance, \citet{parry2021systematic} show users' self-reports can underestimate or overestimate their media use, 
\citet{ernala2020well} find that Facebook users overestimate their usage time and underestimate their number of visits,
and \citet{goetzen2023likes} show that TikTok users tend to overestimate their usage time.

Researchers have recommended using specific wording and multiple-choice questions to avoid the discrepancies~\cite{boase2013measuring,scharkow2016accuracy,ernala2020well}.
Alternatively, existing research also suggests combining surveys and log data can be an appropriate and timely research method~\cite{jurgens2020two, 10.1016/j.chb.2023.108014}.

Our work falls under the method of combining survey responses and digital trace data.
However, we do not rely on participants' responses to \emph{objective questions}, where their log data provides a kind of objective ground truth. However, we rely on self-reports for their \emph{subjective feelings} towards their TikTok usage -- something that is needed to identify addiction, but something log data cannot provide. 

\subsection{Behavioral Addiction on Social Media}
\label{subsec:related-addiction}
Behavioral addiction in online social media has gained traction in recent years~\cite{doi:10.2466/02.09.18.PR0.110.2.501-517, tomczyk2023real}.
One line of prior work is aimed at quantifying the degree of addiction~\cite{doi:10.2466/02.09.18.PR0.110.2.501-517, wolniczak2013association, turel2012benefits}.
In particular, \citep{doi:10.2466/02.09.18.PR0.110.2.501-517} developed the Bergen Facebook Addiction Scale (BFAS) that aims to measure Facebook addiction according to the following six addiction criteria: salience, relapse, mood modification, conflict, withdrawal, and tolerance.
Each item is scored on a five-point scale. 
The authors have suggested a polythetic scoring: a person is considered to be addicted if scoring 3 or above on at least four of the six items. 
Later BFAS has been further developed to measure addiction on social networks more broadly~\cite{balcerowska2020meaningful}. 
BFAS has been shown to be in line with diagnostic addiction criteria ~\cite{andreassen2015online, american2013diagnostic} and has good psychometric properties~\cite{doi:10.2466/02.09.18.PR0.110.2.501-517,andreassen2013relationships}.
Our work measures TikTok addiction using a similar questionnaire.

The other line of work tries to identify different phenomena and how they may end up affecting users' addiction to social media.

For instance, \citep{afacan2019investigation} find that there is a significant relationship between internet usage time and addiction among high school students,
\citep{tomczyk2023real} show that problematic smartphone use weakly correlates with screen time, and
\citep{blackwell2017extraversion} measure participants' levels of extraversion, neuroticism, attachment styles, and fear of missing out and find that these factors are predictive of their addiction to social media.
Furthermore, drawing on gratifications theory, \citep{xu2012not,gan2018gratifications} show that users' staying on social media platforms is affected by hedonic gratifications (e.g., passion and excitement).

In different parts of the current work, we take motivation from a number of the aforementioned work to understand the different usage patterns of participants belonging to different addiction groups.
We shall elaborate on each of those in their corresponding contexts while trying to answer RQ2.

\section{RQ1: Do participants suffer from behavioral addiction?}
\label{sec:survey}


\paragraph{Participant recruitment}

We used Prolific\footnote{\url{https://www.prolific.com/}} to recruit participants, targeting users who regularly use TikTok from the United States, the United Kingdom, and the European Economic Area.
‌In order to ensure the recruitment of high-quality participants, we adopted several screening settings.
Specifically, we required each participant to be fluent in English and to have at least 50 previous submissions with an approval rate of at least 95\%.
Based on our design choices and screening criteria, the number of eligible participants across the aforementioned countries was \num{19698}.
We released the survey for participation to a gender-balanced sample of Prolific users and recruited \num{1590} participants to respond to our questionnaire.
The median time for participants was 3 minutes, and compensation was \textsterling 0.45 (equivalently \textsterling 9 per hour, which is a recommended rate by Prolific\footnote{\url{https://researcher-help.prolific.com/hc/en-gb/articles/360009223533-What-is-your-pricing}}).

\paragraph{Questionnaire}
The survey consists of 28 questions that are grouped into two sections: 1)~demographics~(12 questions) and 2)~TikTok usage habits~(16 questions).
Example questions from each of the sections can be found in \Cref{app:question}.
In the demographics questions, participants are asked to provide among other things their gender, age group, countries, and states/provinces (which helps us determine their time zones). 

The core of our usage questions consists of an adaptation of the Bergen Facebook Addiction Scale (BFAS)~\cite{doi:10.2466/02.09.18.PR0.110.2.501-517} into the context of TikTok (corresponding questions can be found in \Cref{app:BFAS}).
BFAS was initially developed to study Facebook addiction based on the following criteria: salience, tolerance, mood modification, relapse, withdrawal, and conflict. 
All of the six items have been shown to be in line with diagnostic addiction criteria~\cite{andreassen2015online,american2013diagnostic,world1992icd}.

Each question is measured on a five-point scale from \textit{very rarely} to \textit{very often}, scoring 1-5, respectively. 
For convenience, we define the \textbf{addiction score} for a participant to be the total number of questions for which they have selected a score of 3 or above.
Moreover, we explicitly ask participants if they agree with the statement ``I consider myself addicted to TikTok", using a five-point Likert scale question with options from \textit{strongly disagree} to \textit{strongly agree}.

\paragraph{Quality checks}
We calculate Cronbach's $\alpha$~\cite{cronbach1951coefficient} on the six addiction questions, giving a score of 0.82, indicating a good internal consistency of adapted questions in the present study according to standard interpretation~\cite{tavakol2011making}. 
As an additional quality control, we wanted to check if higher (or lower) levels of apparent TikTok addiction could be false positives due to (likely) junk survey responses. 
To test this, we coded a set of 11 implausible answer combinations, e.g. a person is aged between 18 to 24 but also widowed (see \Cref{app:implausible} for full list). 
A user is labeled as implausible if they have at least one such answer.
Overall, 0.9\% of users are labeled as implausible: 1.8\%, 0.4\%, and 0.7\% from HLA, MLA, and LLA, respectively.
We conduct a $\chi^2$ test with a null hypothesis that there is no significant difference regarding addiction distribution between plausible and implausible users.

If the distributions are similar, then we are assured that implausibility is not associated with any addiction group.
The resulting p-value is 0.07, indicating that there is weak to no statistical evidence supporting that implausibility is linked to any addiction group, i.e., that it is unlikely that different levels of addiction are due to implausible responses.
Even if HLA users were more likely to be implausible, we did not find any evidence for implausibility among the subset of users who donated their data (\Cref{sec:donation}).

\paragraph{Assigning addiction levels}
Unlike existing studies where participants are classified as addicted or not~\cite{doi:10.2466/02.09.18.PR0.110.2.501-517}, we stratify participants into three different addiction levels based on their survey responses to have a more fine-grained perspective.
A participant is said to be \textit{highly likely addicted} (HLA) if they have an addiction score $\ge4$, or alternatively, responded \textit{agree} or \textit{strongly agree} to the explicit addiction question.
The threshold used is aligned with existing literature~\cite{doi:10.2466/02.09.18.PR0.110.2.501-517}. 
A participant is said to be \textit{moderately} or \textit{less likely addicted} (MLA or LLA) if they have an addiction score of 2 or 3, or less than 2, respectively\footnote{Alternatively, we have experimented with different thresholds: 1, 2, and 3. We have qualitatively similar results, thus we stick to a score of 2, which is also the median of participants' addiction scores.}.

\begin{table}[h]
    \centering
    \small
    \setlength{\tabcolsep}{3pt}
    \begin{tabular}{c c cc cc cccc}
    \toprule
    & \multicolumn{2}{c}{LLA} & \multicolumn{2}{c}{MLA} & \multicolumn{4}{c}{HLA} \\ 
    \cmidrule(lr){1-1} \cmidrule(lr){2-3}  \cmidrule(lr){4-5}  \cmidrule(lr){6-8} \cmidrule(lr){9-9}
    Addiction Score & 0 & 1 & 2 & 3 & 4 & 5 & 6 & + Explicit \\
    \# Participants & 306 & 367 & 292 & 267 & 180 & 107 & 71 & 276 \\
    \cmidrule(lr){1-1} \cmidrule(lr){2-3}  \cmidrule(lr){4-5}  \cmidrule(lr){6-9} 
    Total & \multicolumn{2}{c}{667} & \multicolumn{2}{c}{487} & \multicolumn{4}{c}{436} \\

    \bottomrule
    \end{tabular}
    \caption{TikTok addiction level breakdown of the recruited participants based on addiction scores. In addition, \textit{Explicit} represents the number of participants answering \textit{agree} or \textit{strongly agree} to the explicit addiction question.}
    \label{tab:addition-breakdown}
\end{table}

\Cref{tab:addition-breakdown} reports the participants by their addiction classification.
There are 436 HLA participants in total, of which 358 are classified as HLA by BFAS and 276 are classified as HLA by the explicit question, with an overlap of 187. 
Moreover, we notice 39\% of participants from the age group 18-24 belong to this category (see \Cref{app:demo-survey} for a detailed breakdown).
The remaining \num{1154} participants constitute \num{487} MLA and \num{667} LLA participants.



\begin{figure}[ht]
    \centering
    \begin{subfigure}[t]{0.45\columnwidth}
        \centering
        \includegraphics[width=1\columnwidth]{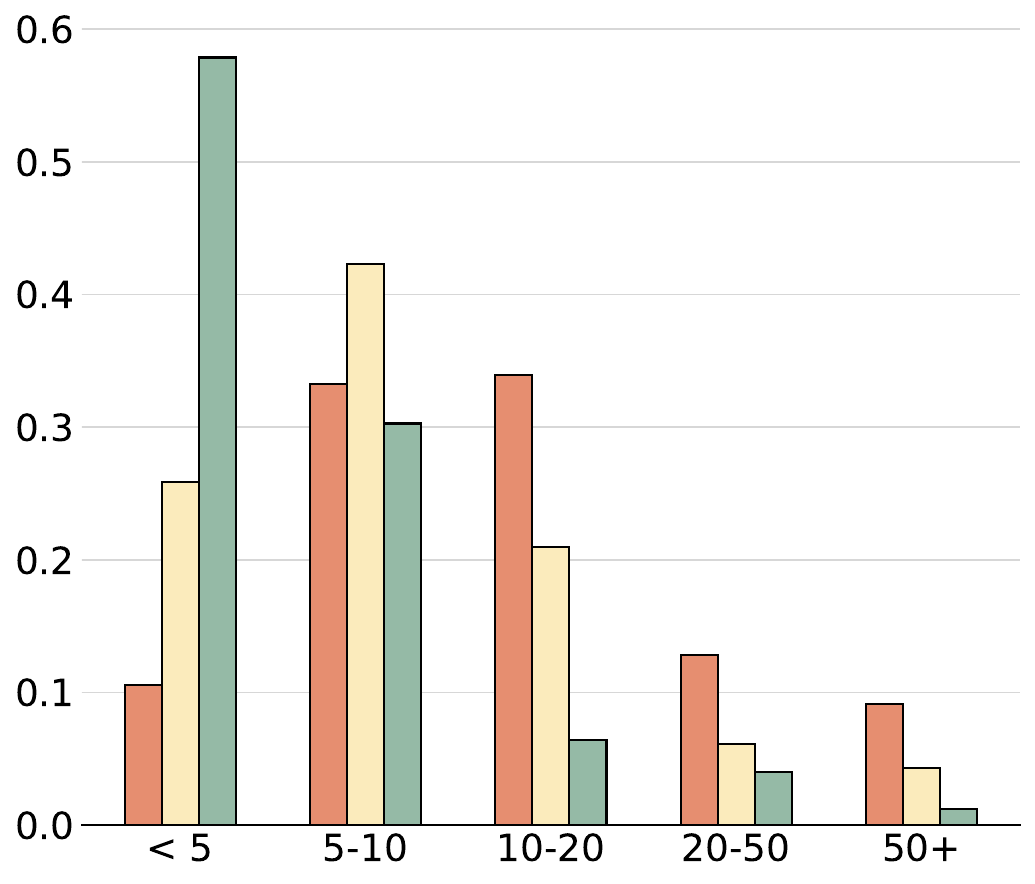}
        \caption{How often do you use TikTok per day?}
    \end{subfigure}
    \begin{subfigure}[t]{0.47\columnwidth}
        \centering
        \includegraphics[width=1\columnwidth]{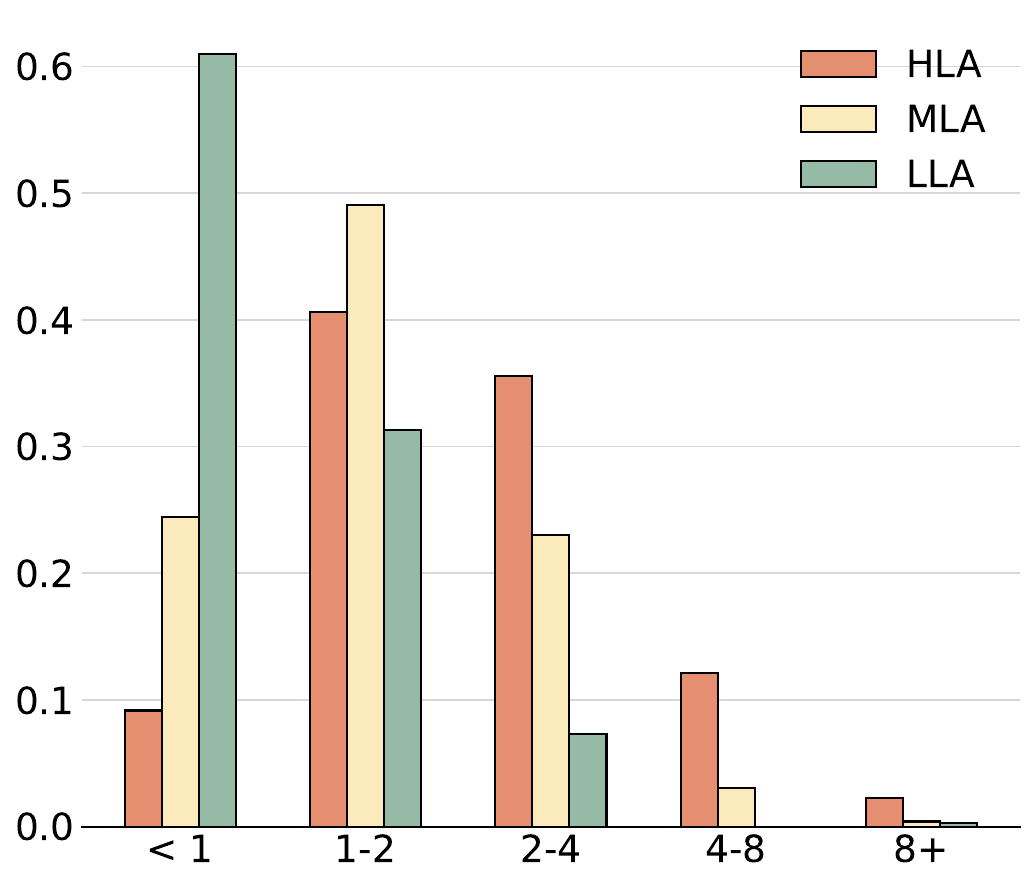}
        \caption{How many hours per day do you spend on TikTok?}
    \end{subfigure}
    
    \caption{Participants' answers for TikTok usage questions across different addiction levels. The usage pattern of highly likely addicted (HLA) and less likely addicted (LLA) participants are distinct: HLA participants report to spend more time on TikTok and use it more often than LLA participants.}
    \label{fig:usage-question-vs-addict}
\end{figure}

Motivated by the fact that 276 out of the 436 HLA participants responded positively (either \textit{agree} or \textit{strongly agree}) to the explicit addiction question, we assess how participants answered this question based on their addiction scores.
We find that less than 1\% of participants with addiction scores of 0 and 1 respond positively to the explicit addiction question.
This percentage is 13\% for addiction scores of 2 and 3, and it goes up to 55\% for participants with addiction scores of 4 or above.

Meanwhile, we also observe that 71\% of participants who self-report as being addicted have an addiction score of 4 or above.
A detailed breakdown of participants' answers by addiction group can be found in \Cref{app:explicit-question}.
In summary, we find that participants with higher BFAS-based addiction scores are more likely to self-report being addicted -- and vice versa -- indicating consistency between the results of the psychological method and the participants' perception of their behavioral addiction level.
Based on this consistency, and to have higher coverage of HLA users, we include users with self-reported addiction into the HLA group.
There are 78 users included in the HLA group (18\%) solely due to their positive response to the explicit question.

Next, we investigate how participants from different addiction groups responded to ``TikTok usage habits'' questions.
\Cref{fig:usage-question-vs-addict} displays how users from each addiction group use TikTok as per their own survey responses to two ``TikTok usage habits'' questions.
When asked about the frequency of TikTok usage, nearly 60\% of LLA users responded they use TikTok less than 5 times, while the most selected options for the MLA and HLA groups are 5-10 and 10-20 times, respectively.
We have observed similar differences in the selected options for the amount of time spent on TikTok.
The responses here indicate that the addiction level is associated with usage intensity, which is not a priori obvious, as addiction is defined through the harm the usage causes, not through the quantity of usage.

\paragraph{Main Takeaways of Section}
\begin{enumerate}
    \item End users may potentially suffer from TikTok addiction, as 27\% of all participants are classified as highly likely addicted. 
    \item Addiction level correlates to usage intensity as per participants' survey responses.
\end{enumerate}

\section{Data Donation}
\label{sec:donation}

The crowdsourced survey brings out the existence of behavioral addiction among TikTok users.
To further understand this phenomenon, we need to understand how they engage with different content and features on TikTok.
To this end, we invite all the \num{1590} survey participants to donate their TikTok data by exercising their GDPR right of access by data subjects.
To facilitate the process, we develop a data donation platform where we first show participants how they can request their data on TikTok mobile apps (see \Cref{app:instruction}). In the second phase, they are required to come back to the platform and donate the data that they received from TikTok upon their consent. 

Following prior work, we keep donating video browsing history as a mandatory field and the participants are remunerated with a reward of \$5 for the same.
None of the other fields in the TikTok data (e.g., search history, comment, ads, etc.) are mandatory for donation; however, participants are welcome to donate these if they wish to.
Donation of each additional field is compensated by \$1, and the maximum remuneration per participant is \$16, which is paid to participants through Prolific\footnote{In our case, Prolific works with GBP, and we convert USD to GBP when processing the payment}.
The donation platform also removes sensitive personal information, such as private messages and IP addresses, before accepting the participant's data.

TikTok only provides video browsing history of the last 180 days of an account.
In order to ensure receiving high-quality data from the participants, we only accept video browsing histories consisting of at least 90 unique days. 
With a threshold of 90 days, we can have users who are active for at least half of the maximum time window.\footnote{ We have experimented how our observations change when we increase the threshold to 100/120/150 days, and have qualitatively similar findings.} 
Ineligible participants will be disallowed from uploading their data and only be reimbursed for screening by \textsterling 1.


\begin{table}[ht]
    \centering
    \small
    \begin{tabular}{crccc}
    \toprule
           &  & LLA & MLA & HLA \\
    \midrule
    \multirow{3}{*}{Gender} & Female & 12 & 18 & 24\\
    & Male & 21 & 14 & 14\\
    & Non-binary & 2 & 1 & 1\\
    \midrule
    \multirow{4}{*}{Age} & 18-24 & 7 & 5 & 17 \\
    & 25-34  & 20 & 19 & 13\\
    & 35-44 & 5 & 6 & 6\\
    & 45-64  & 3 & 3 & 3\\
    \midrule
    \multirow{3}{*}{Mobile OS} & iOS & 22 & 22 & 28 \\
    & Android & 13 & 11 & 10 \\
    & Prefer not to say & 0 & 0 & 1 \\
    \bottomrule
    \end{tabular}
    \caption{Gender, age, and mobile operating system distribution within the three addiction groups of the participants who donated their TikTok data. }
    \label{tab:demo-from-data}
\end{table}

\begin{figure}[t]
  \centering
  \includegraphics[width=0.9\columnwidth]{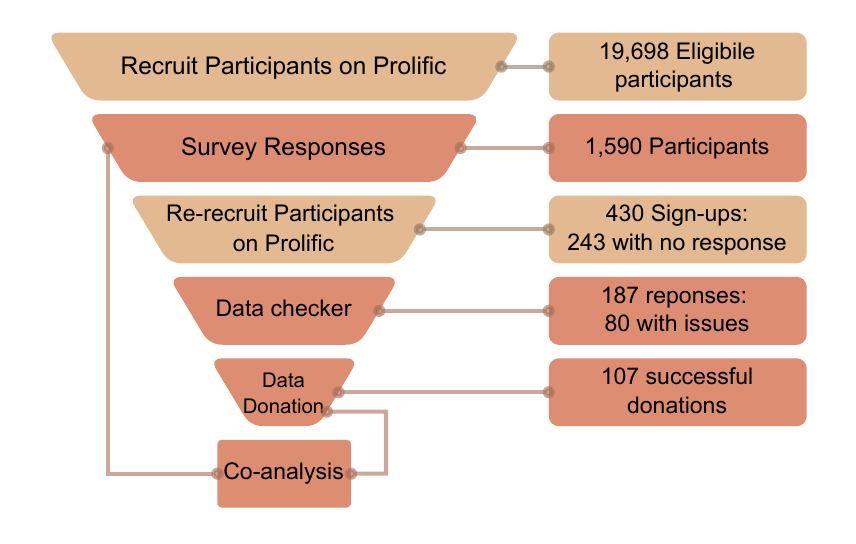}
  \caption{Data collection pipeline and the number of participants at each stage.}
  \label{fig:diagram}
\end{figure}

The pipeline of our data collection is shown in \Cref{fig:diagram}.
We released our study to all the \num{1590} participants who filled out our survey.
Among these participants, 430 signed up in the data donation phase, but 203 of them did not get back to us, most likely due to privacy-related concerns.
There were 187 participants who got back to us, among whom 48 had less than 90 days in their browsing histories, and 32 had mismatched formats (e.g., requesting a TXT file rather than JSON). 
Our system does not allow such participants to proceed.
In the end, we received donations from 107 participants, 92 of whom were from the United States and 15 from Europe.
Among these 107 participants, 74 of them decided to donate all the available additional fields.
Also, these 107 participants are divided into 39 HLA, 35 MLA, and 33 LLA.
In total, our collected dataset contains \num{4231317} TikTok videos viewing instances from the video browsing histories, among which there are \num{2769657} unique videos.
We convert all the timestamps of activities, such as video browsing and engagement, into each user's local time zone based on their self-reported country and state/province.

\Cref{tab:demo-from-data} presents the demographics in each addiction group.
We notice an imbalanced distribution across groups.
The HLA group contains predominantly female participants (62\%), while the LLA group contains mostly male participants (60\%).
Meanwhile, the HLA group consists of many young adults, whereas MLA and HLA have more middle-aged participants.
In addition, more than half of the participants use TikTok on iOS, with the difference between operating systems most pronounced in the HLA group.

\paragraph{Video metadata collection}
The video browsing history provided by TikTok only contains the URL to the video and the timestamp at which the user started viewing it.
We further collected each video's metadata by using a modified version of the unofficial Python API~\cite{tiktokapi}.
The collected metadata contains each video's description, duration, engagement statistics (e.g., number of views, likes, shares, etc.), and author information.
Overall, we successfully obtained metadata for 88\% of all unique videos from our dataset. The remaining videos were either removed from TikTok or the TikTok account that posted them had changed their account to private at the time of metadata collection.

\paragraph{Inferring viewing duration and session}
As previously stated, the video browsing history does not contain information about how much time a user spends (partially) on watching the video.
Since this piece of information is crucial for the purposes of our study, we infer it based on the timestamps of viewing videos~\cite{zannettou2024analyzing}.
We define the inferred viewing duration of a video as the gap between the timestamp it was viewed and the timestamp for the next video in the video browsing history.
The inferred viewing duration reflects the actual time a user spent watching a video.
From now on, we simply refer to inferred viewing duration as ``viewing duration''.

Another important ingredient for the purposes of our study is the notion of ``session''.
We define a session to be an uninterrupted viewing of TikTok videos.
We utilize the methodology proposed by \citet{halfaker2015user} and apply a two-component Gaussian mixture model to users' viewing duration.

The idea is that one component (within-session) captures the actual duration spent on watching videos, while the other component (between-session) captures user breaks.
We can identify the threshold by finding the point where viewing duration is equally likely under both components.
A histogram of users' viewing duration and fitted mixture models can be found in \cref{app:halfaker-viewing}.
We derive a session break threshold of 181 seconds, meaning that videos with viewing durations higher than this threshold indicate a likely break in the TikTok viewing session.
Hence, we remove such videos and split each user's video viewing timeline into sessions. 
We find this threshold reasonable since 96\% of all viewing duration is at most 181 seconds.

Using this threshold, for 96 users (89.7\% of all users), we keep more than 90\% of each individual's videos watched.
For 103 users (96.2\% of all users), we keep more than 80\% of each individual's videos.
We choose not to remove videos with viewing durations higher than the threshold but less than the actual duration (if it exists). 
This is to avoid removing videos with actual duration longer than the threshold, where it is normal for users to spend higher viewing durations.
There are 0.14\% of all videos falling into this category.
Overall, we keep 96.1\% of all the videos. 

\paragraph{Completeness checks for logged data}

For the log data, we ultimately rely on TikTok's black box logging algorithm to work correctly. 
However, such logging pipelines can fail and Meta had to admit that data they had released under the Social Science One initiative were incomplete~\cite{metadata}. 
So to check the completeness of the logged data, the authors did some spot checks, comparing their actual interaction on TikTok with what the log data contained. 
In all of the cases, the log data was complete. Details are in \Cref{app:data-complete}.

\section{RQ2: Do participants in different addiction groups exhibit different usage patterns?}
\label{sec:usage-content}

In this section, we analyze the donated TikTok dataset to characterize differences between addiction groups based on their TikTok usage and content patterns.
Existing work mostly relies on surveys and is limited to a small set of concepts they can examine.
With the help of actual data, we can study addiction from a multifaceted perspective.
Particularly, we compare addiction groups based on several measures, such as viewing duration, sessions, video engagement, and topics and sentiments of videos.
We present our methodology and findings below.

\subsection{How much time do participants of different addiction groups spend on TikTok?}

Existing studies on addiction~\cite{afacan2019investigation,tomczyk2023real} have pointed at correlation between screen time and internet/social media addiction. 
Motivated by those findings, we hypothesize that more addicted participants have a higher viewing time on TikTok.

\begin{figure}[t]
  \centering
  \includegraphics[width=0.7\columnwidth]{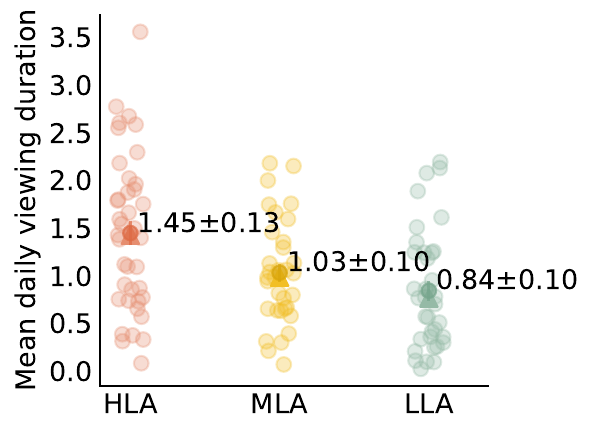}
  \caption{Average daily hours spent on watching videos on TikTok. Circular dots represent mean values and vertical bars represent standard errors. Triangular dots represent median values, same for later figures. HLA users spend more time on TikTok watching videos than LLA users.}
  \label{fig:viewing}
\end{figure}

We first calculate each user's daily viewing time by summing over the viewing duration of each video each day.
We then compute the mean daily viewing time for each user and then take the average within each addiction group.
\Cref{fig:viewing} displays the mean daily time (in hours) spent watching TikTok videos for each of the addiction groups.
The mean daily watching times for HLA, MLA, and LLA are 1 hour and 27 minutes, 1 hour and 2 minutes, and 50 minutes, respectively.

We perform a one-sided permutation test~\cite{efron1994introduction,ernst2004permutation} on the differences in mean viewing time across groups with \num{10000} resamples (we apply the same test in later comparisons; thus, the setup is omitted for simplicity).
We chose the permutation test as it is non-parametric and our sample size is relatively small.
The test results confirm that HLA users have a higher average viewing hours than MLA and LLA users (HLA and MLA: $p<0.01$, Cohen's $d=0.86$; HLA and LLA: $p<0.01$, Cohen's $d=0.83$); thus confirming our hypothesis. 
However, no significance is found between the other pairs of groups.

\subsection{How do participants of different addiction groups spend time on TikTok?}

\begin{figure*}[ht]
    \centering
    \begin{subfigure}[t]{0.66\columnwidth}
        \centering
        \includegraphics[width=1\columnwidth]{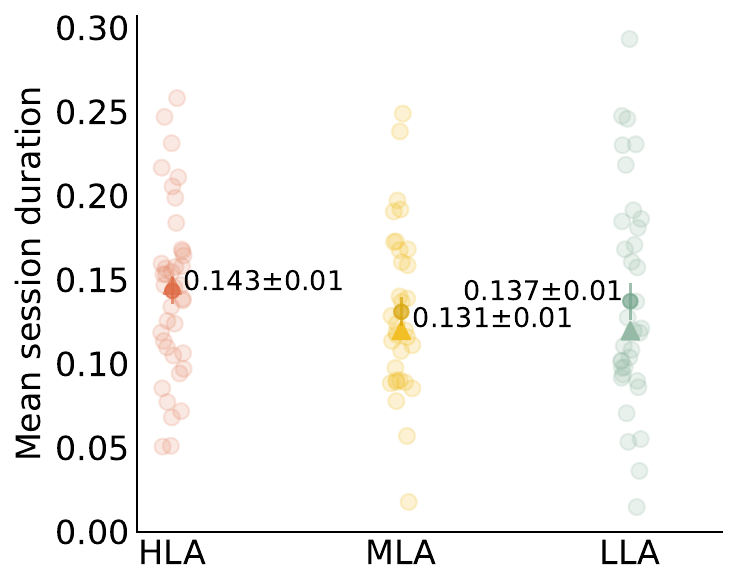}
        \caption{Session duration}
        \label{subfig:sess-dur}
    \end{subfigure}%
    ~
    \begin{subfigure}[t]{0.66\columnwidth}
        \centering
        \includegraphics[width=1\columnwidth]{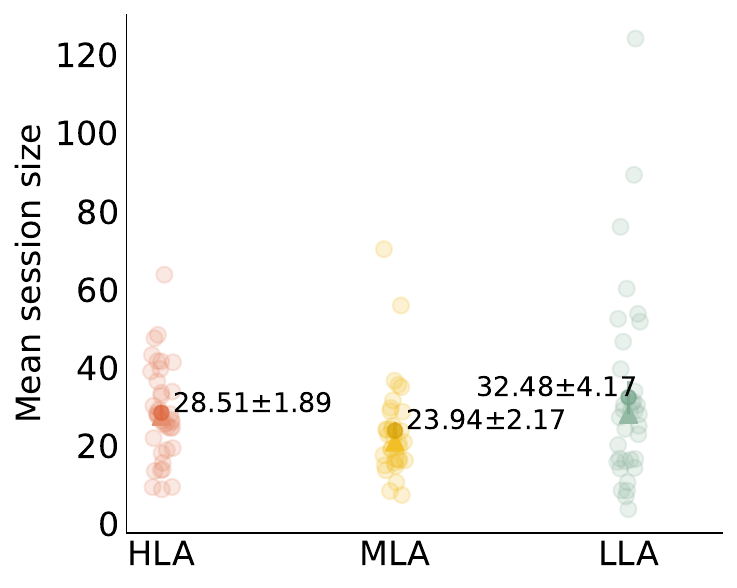}
        \caption{Session size}
        \label{subfig:sess-size}
    \end{subfigure}
    ~
    \begin{subfigure}[t]{0.66\columnwidth}
        \centering
        \includegraphics[width=1\columnwidth]{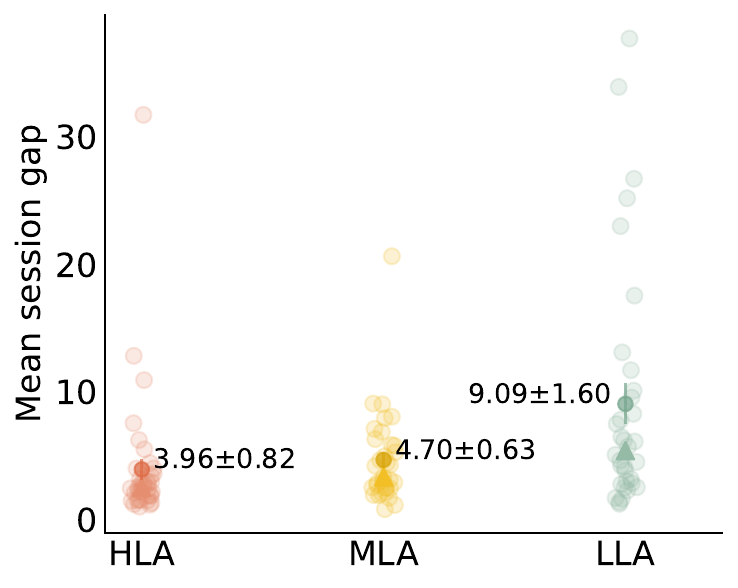}
        \caption{Session gap}
        \label{subfig:sess-gap}
    \end{subfigure}
    
    \caption{Session measures across addiction groups. Session duration is measured in hours. HLA and MLA users return to TikTok more often than LLA users and MLA users also have watched fewer videos each time than LLA users.}
    \label{fig:sess}
\end{figure*}

As a follow-up question to the average viewing time, here, we ask how the time they spend is distributed within a day.
We hypothesize that addicted users have longer viewing sessions where they see more videos and come back to the platform more frequently.
We consider several session related measures to help us answer this question.
We define session duration as the time span of a session, session size as the number of videos within a session, and session gap as the temporal gap between two consecutive sessions.

\Cref{fig:sess} depicts the mean of different session measures across the three addiction groups.
We can see that increasing addiction does not imply a monotonic change in session measures. 
\Cref{subfig:sess-dur} displays the mean session duration (in hours) for each group.
Most users have short sessions, with nearly all of them having at least half their sessions shorter than 10 minutes. 
The mean session duration for HLA, MLA and LLA users are 8.58, 7.86 and 8.10 minutes respectively.
We perform a permutation test and do not observe any difference across groups.

\Cref{subfig:sess-size} shows the number of videos watched with a session.
We find MLA users watch significantly less videos than LLA users ($p<0.05$, Cohen's $d=0.2$).
\Cref{subfig:sess-gap} displays mean session gaps (in hours) for each group. 
HLA and MLA users return to TikTok on average every 4 to 5 hours, while LLA users return every 9 hours.
Both HLA and MLA users have smaller mean session gaps than LLA users, as shown by the permutation test (HLA and LLA: $p<0.01$, Cohen's $d=0.69$; MLA and LLA: $p<0.01$, Cohen's $d=0.72$). 
The results suggest that, despite spending similar time on TikTok per session, HLA and MLA users return to TikTok more often than LLA users.

Furthermore, we examine the time of the day that users in the three addiction groups use TikTok.
For simplicity, we check the percentage of watching time per day spent from 6:00 am to 6:00 pm as a proxy for measuring how users use TikTok during the daytime.
We take the average over time for each user and then within each group.
The average percentage of time spent during daytime for HLA, MLA and LLA users is $0.44 \pm 0.16$, $0.50 \pm 0.14$ and $0.51 \pm 0.18$, respectively.
HLA users spend statistically less time during daytime on TikTok, and more at night, compared to MLA and LLA users (HLA and MLA: $p<0.05$, Cohen's $d=0.46$; HLA and LLA: $p<0.05$, Cohen's $d=0.42$).

If user classification is (partly) based on usage time, then, trivially, this may lead to the observed differences in metrics related to time across groups.
Importantly, the BFAS questions used to classify users' addiction levels do not directly involve usage time.
However, some questions (e.g., Conflict) might be suspected of having an indirect relationship with usage time, potentially influencing our observations.
To demonstrate this indirect relationship does not significantly impact our findings, we conduct a set of experiments by manually resetting responses to these questions.
Our findings have shown that the overall observations remain consistent, supporting the robustness of our results.
Further details on the experiment can be found in \Cref{app:question-check}.

\subsection{How do participants of different addiction groups engage with videos on TikTok?}

\begin{figure}[t!]
    \centering
    \begin{subfigure}[t]{0.48\columnwidth}
        \centering
        \includegraphics[width=1\columnwidth]{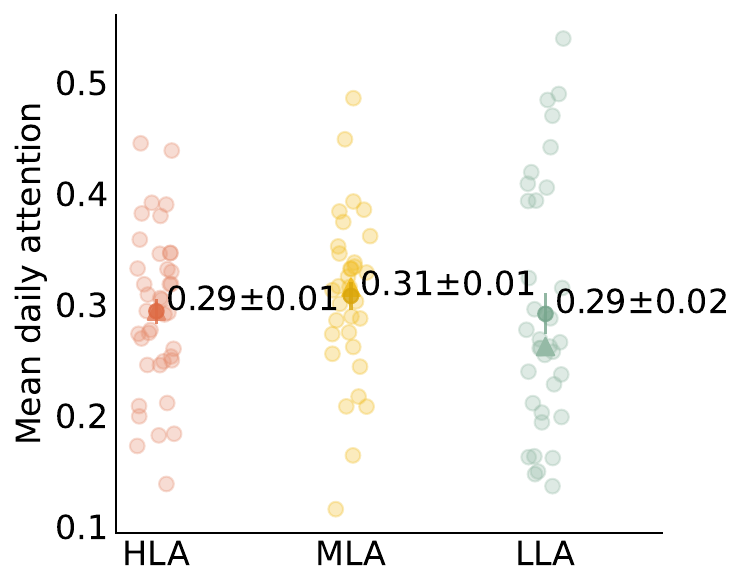}
        \caption{Video attention (passive)}
        \label{subfig:videos-engage-attention}
    \end{subfigure}%
    ~
    \begin{subfigure}[t]{0.52\columnwidth}
        \centering
        \includegraphics[width=0.95\columnwidth]{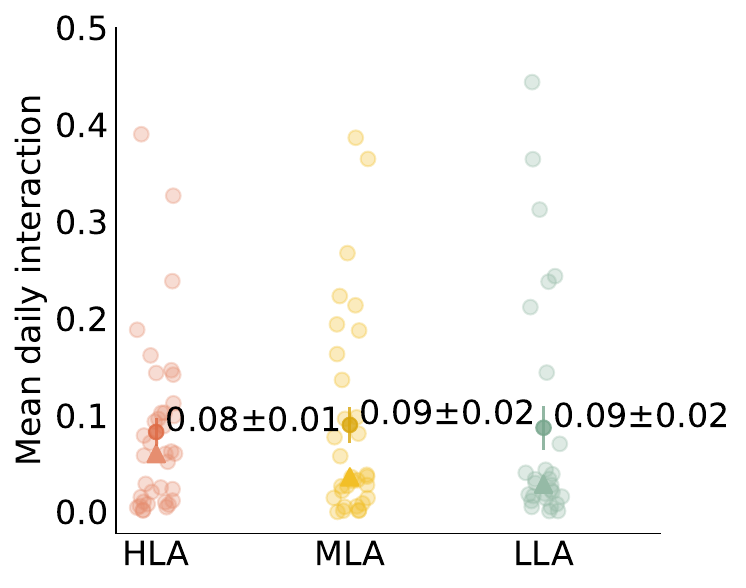}
        \caption{Video interaction (active)}
        \label{subfig:videos-engage-interaction}
    \end{subfigure}
    
    \caption{Mean daily video engagement. No difference is observed in users' attention and interactions with videos across addiction groups.}
    \label{fig:videos-engage}
\end{figure}

Prior works~\cite{verduyn2021impact,valkenburg2022associations} show contradictory observations on the relation between passive social media use and mental well-being.
Examples of passive use include scrolling through one's feeds without direct engagement.
In this work, we look into both passive and active engagement on TikTok.
We hypothesize that more addicted users have more passive and less active engagement on TikTok. 

To understand this phenomenon, we consider two engagement signals: 1) attention and 2) interaction~\cite{zannettou2024analyzing}.
``Attention'', a passive engagement, refers to watching a video until the end, whereas ``interaction'', an active engagement, refers to liking, sharing, or favoring a video. 
We do not consider comments since the users' downloaded TikTok data does not contain the video ID for a comment.

We measure average daily engagement with TikTok videos by finding the mean daily attention and interaction for each user.
Mean daily attention is defined as the average fraction of videos watched until the end per day, whereas mean daily interaction is defined as the average fraction of videos that the user liked, shared, or favored per day.
After finding the mean daily attention and interaction of all users, we take the mean within each addiction group.
\Cref{fig:videos-engage} displays users' mean daily attention and interaction with videos they have viewed across the three addiction groups.
Particularly, according to \Cref{subfig:videos-engage-attention}, we see that the mean daily video attention across HLA, MLA, and LLA groups are \num{0.29}, \num{0.31}, and \num{0.29}, respectively.
\Cref{subfig:videos-engage-interaction} shows that the mean daily video interaction across HLA, MLA, and LLA groups are \num{0.08}, \num{0.09}, and \num{0.09}, respectively.

Active engagement takes place less often compared to passive engagement, as it is a more involved and explicit social media behavior~\cite{muntinga2011introducing}.
While the passive engagement is similar across the participants of different addiction groups, we observe a notable relative difference between the active engagement of HLA and others.
However, we do not observe any statistical difference across groups based on the permutation tests.



\subsection{How do the authors of provided / consumed videos differ across participants of different addiction groups?}

Fear of missing out~\cite{przybylski2013motivational} refers to one's feeling about missing from what the peers are doing, and it has been shown to be associated with addiction 
~\cite{blackwell2017extraversion,tunc2019smartphone}.
Although users do not necessarily follow their peers on TikTok, investigating how they engage with videos from their followings may still reveal some insights into how different participants engage with content from different authors.
We hypothesize that more addicted users engage with more videos from their followings than others.

\begin{figure}[t!]
    \centering
    \begin{subfigure}[t]{0.49\columnwidth}
        \centering
        \includegraphics[width=1\columnwidth]{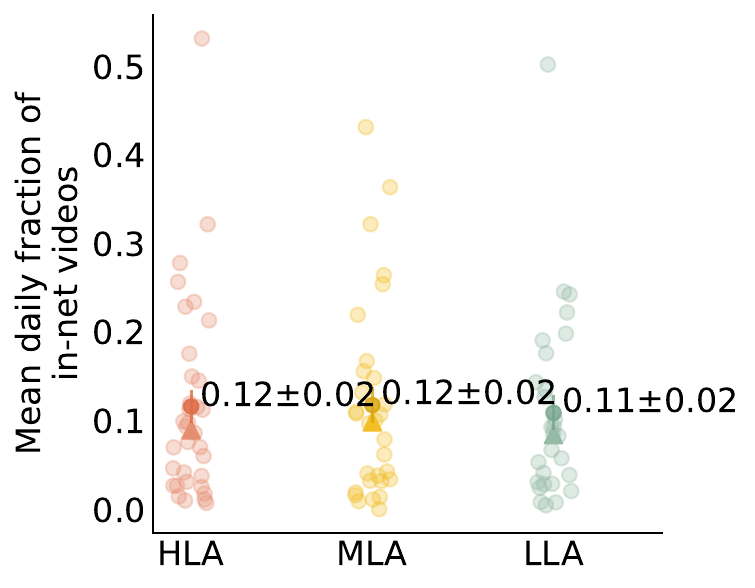}
        \caption{Provision from TikTok}
        \label{subfig:innet-provision}
    \end{subfigure}
    \begin{subfigure}[t]{0.49\columnwidth}
        \centering
        \includegraphics[width=1\columnwidth]{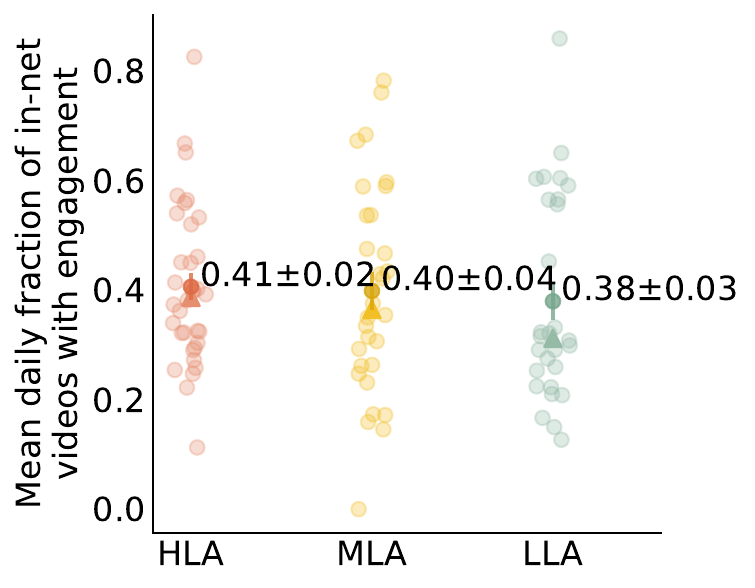}
        \caption{Engagement by users}
        \label{subfig:innet-engage}
    \end{subfigure}
    
    \caption{Mean daily fraction of in-network videos and engagement with them. No difference is observed in the provision and engagement of in-network videos.}
    \label{fig:videos-seeking}
\end{figure}

We check the authors of each video against the user's following list to determine whether the videos are from their followings.
We refer to this set of videos as in-network videos.
\Cref{subfig:innet-provision} shows the provision of in-network videos to users in different addiction groups. In general, the provision rates are observed to be similar across the board.
The in-network video ratio remains low among users, with 75\% of all users having watched less than 15\% in-network videos.
One potential reason is that the ``For-you-page'' is the default feed on TikTok, and users usually do not actively change that setting but rather stick to the algorithmically curated content.

Next, we compute the fraction of in-network videos that users of different addiction groups actively and passively engaged with. 
\Cref{subfig:innet-engage} shows the mean daily fraction of in-network videos with engagement across the three addiction groups.
The fraction across HLA, MLA, and LLA groups are \num{0.41}, \num{0.40}, and \num{0.38}, respectively.
We do not observe any difference across groups based on the permutation tests.
We also repeat the calculation for out-network videos, i.e., videos that are outside each user's following list (HLA: 0.34; MLA: 0.35, LLA: 0.34).
We find that, on average, each group has a higher engagement ratio on in-network videos than on out-network videos.
The difference between in-network and out-network engagement for HLA, MLA, and LLA users is 0.07, 0.05, and 0.04 respectively.

\subsection{How do sentiments of consumed videos differ across participants in addiction groups?}

Existing research on uses and gratifications theory has demonstrated that hedonic gratifications affect users' use and continuance on social media platforms~\cite{xu2012not,gan2018gratifications}.
Watching positive videos, as one type of hedonic gratification, may contribute to users' stay on TikTok.
For this reason, we hypothesize that more addicted users may tend to watch more positive videos.

We detect the sentiment of each video by applying XLM-T~\cite{barbieri-espinosaanke-camachocollados:2022:LREC} to the video's description. 
XLM-T is a multilingual model developed for sentiment analysis on Twitter. 
It assigns probability scores to three possible labels: positive, neutral, and negative, and returns the label with the highest probability.
To validate its performance on TikTok video descriptions, three co-authors independently annotated a random set of 100 video descriptions.
The agreement between majority-voted annotations and model output is 73\%, which ensures the reliability of XLM-T in the context of TikTok.
It should be noted that detecting sentiments using video content is out of the scope of this work, and we focus on texts as they are straightforward to work with.
Existing work has also demonstrated the effectiveness of using TikTok video descriptions~\cite{10.1145/3485447.3512102}.
For simplicity, we refer to videos with positive (negative) sentiments as positive (negative) videos.
Some examples of video descriptions with positive sentiments can be found in \Cref{app:positive-example}.

\begin{figure}[t!]
    \centering
    \begin{subfigure}[t]{0.49\columnwidth}
        \centering
        \includegraphics[width=1\columnwidth]{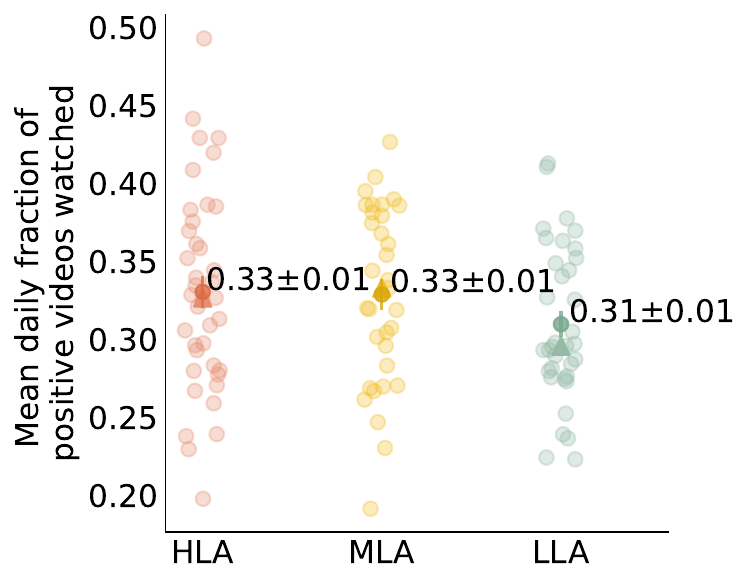}
        \caption{Provision from TikTok}
        \label{subfig:pos-provision}
    \end{subfigure}%
    ~
    \begin{subfigure}[t]{0.49\columnwidth}
        \centering
        \includegraphics[width=1\columnwidth]{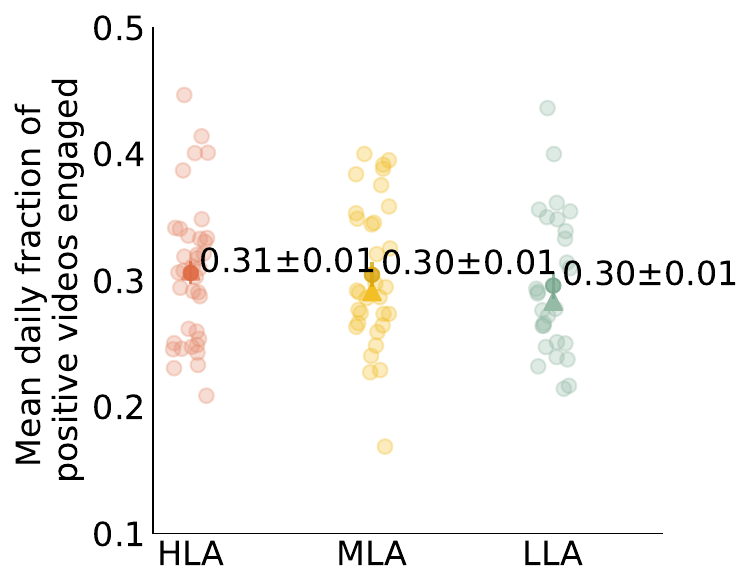}
        \caption{Engagement by users}
        \label{subfig:pos-engage}
    \end{subfigure}
    \caption{Mean daily fraction of videos with positive sentiment and their engagement. No difference is observed across groups. The engagement with positive videos is higher than the engagement with overall videos, as shown in Figure~\ref{fig:videos-engage}.}
    \label{fig:videos-pos-video}
\end{figure}

We find that negative videos take only a small portion of each user's browsing history, ranging from 8\% to 30\%, while positive videos range from 20\% to 49\%.
Next, we measure the mean daily fraction of positive videos across the addiction groups.
To do so, we calculate the mean daily fraction of positive videos for each user and then compute the mean within each addiction group.
\Cref{subfig:pos-provision} shows the computed fraction across the three addiction groups.
The mean fraction for HLA, MLA, and LLA, are \num{0.33}, \num{0.33}, and \num{0.31}, respectively, without any statistically significant differences.

We compute the mean daily fraction of positive videos with engagement over all positive videos in a similar way.
\Cref{subfig:pos-engage} shows the computed fraction.
While we do not observe any statistical difference between the addiction groups, the fractions are higher than those shown in \Cref{fig:videos-engage}, suggesting that users are more actively engaging with positive videos compared to overall videos.

\paragraph{Main Takeaways of Section}

\begin{enumerate}
    \item HLA users spend on average 37 minutes longer per day watching videos than LLA users.
    \item HLA and MLA users on average return to TikTok 2 times more frequently than LLA users.
    \item LLA users watch on average 8 more videos than MLA users per session.
    \item Users with different addiction levels slightly differ with respect to their engagement with in-network videos.
\end{enumerate}

\section{RQ3: Can we predict addiction level based on a user’s social
media data?}
\label{sec:predict}

Previously, we described different usage patterns across the three addiction groups.
In this section, we aim to investigate the feasibility of predicting user's addiction level given their usage patterns on TikTok.
Accurate predictions could help identify addicted people and perform interventions on their social media usage, either on the user side, e.g., through involved usage monitoring apps, or on the platform side, e.g., if platforms were obliged to reduce their systematic risk. 
Notice our goal here is not to develop state-of-the-art models for potential deployment. 
Among other things, the platforms themselves would have additional features at their disposal, such as information about a user's social network, fine-grained details regarding their activity within and beyond the platform of consideration, etc. 
Rather, we demonstrate the feasibility of the task, and investigate which specific features are most useful in this context. 

\paragraph{Classification task: }
We conduct our experiments in two settings: (1)~\textit{multi-label} classification: where the goal is to predict a participant's addiction group, HLA, MLA, or LLA. (2)~\textit{binary} classification: where the goal is to predict if a participant is highly likely addicted or not. 

\paragraph{Features used: }
We select the set of features based on observations from \Cref{sec:usage-content}.
We focus on measures that are shown to be statistically different across addiction groups in our analyses of RQ2.
We end up with the following features: daily viewing time, session size, and session gaps. 
We additionally choose daily positive videos watched as a content-related feature despite marginal significance. 
For each user, all the features are aggregated in the same manner as described in \Cref{sec:usage-content}. 
For brevity, we limit the discussion to features from data donations in the main draft. 
Similar analyses with features from other input sources (e.g., user survey) are explained in Appendix~\ref{app:predict-input} for reference. 

\paragraph{Classifiers used: }
In our experiments, we tested decision tree (DT), logistic regression (LR), support vector machine (SVM), K-nearest neighbors (KNN), and multi-layer perceptron (MLP). 
For all the classifiers, we perform a 10-fold stratified cross-validation.
For each split, perform a grid search on the hyperparameters using the 90\% data and then evaluate models on the remaining 10\% test set. 
In this case, each instance gets evaluated as test data, which helps us better estimate the robustness of the models due to the relatively small sample size.
Both validation and test are evaluated using macro $F_1$ score.
Throughout our experiments, MLP maintains better performance compared to other models.
For simplicity, we report our results on MLP throughout the rest of this section.

\begin{table}[h]
    \centering
    \small
    \begin{tabularx}{0.9\columnwidth}{c rrr rrr}
    \toprule
      \multirow{2}{*}{Class} &  \multicolumn{3}{c}{Multi} & \multicolumn{3}{c}{Binary} \\
         \cmidrule(lr){2-4} \cmidrule(lr){5-7}
        &  P    & R & $F_1$  &  P  & R & $F_1$ \\
    \midrule
     0 (35) & 0.56 & 0.63 & 0.59  & 0.74 & 0.87 & 0.80 \\ 
     1 (33) & 0.50 & 0.33 & 0.40  & 0.67 & 0.46 & 0.55 \\
     2 (39) & 0.54 & 0.64 & 0.59  &   \\
    \midrule 
    Macro Avg. & 0.54 & 0.53 & 0.53 & 0.70 & 0.66 & 0.67 \\
    \bottomrule
    \end{tabularx}
    \caption{Addiction prediction performance. In the multi-label setting, class 0/1/2 represents LLA/MLA/HLA, respectively. In the binary setting, class 0 represents LLA+MLA, while class 1 represents HLA. Numbers within brackets represent class size.}
    \label{tab:predict-model}
\end{table}

\Cref{tab:predict-model} reports the performance of MLP models.
From the recall of the binary setting, we observe that the classifier can identify most non-addicted participants.
However, less than half of HLA users are correctly classified.
The classifier performs relatively better identifying HLA users in the multi-class setting, with a recall of 0.64.
The performance of other models can be found in \Cref{app:predict}.

We consider misclassifications of different addiction levels to be associated with different risks.
Classifying an HLA user as LLA indicates negligence on addiction, which consequently may lead to decreasing mental well-being of the HLA group. 
Whereas a misclassification, on the other way, may lead to potential detection and intervention at an earlier stage. 
The model has achieved a moderate recall on the HLA class under the multi-class setting, demonstrating the potential of using online data to detect some but not all HLA users.
That said, the false-positive rate may prevent such systems from being deployed for practical purposes.

    

\begin{table}[h]
    \centering
    \setlength{\tabcolsep}{5pt}
    \begin{tabular}{lrlr}
    \toprule
     \multicolumn{2}{c}{Multi} & \multicolumn{2}{c}{Binary } \\
     \cmidrule(lr){1-2} \cmidrule(lr){3-4}
    Feature & Drop & Feature & Drop \\
    \cmidrule(lr){1-1} \cmidrule(lr){2-2} \cmidrule(lr){3-3} \cmidrule(lr){4-4}
    Viewing time & 0.13$\pm$0.17  & Viewing time & 0.17$\pm$0.22  \\
    Positive video & 0.06$\pm$0.12  & Positive video & 0.06$\pm$0.11 \\
    Session size & 0.05$\pm$0.10 & Session gap & 0.01$\pm$0.08 \\
    Session gap & 0.03$\pm$0.09  & Session size & 0.01$\pm$0.09 \\
    
    \bottomrule
    \end{tabular}
    \caption{Feature importance evaluated by permutation. Viewing time and positive video contribute the most to the model's generalization of the unseen data under multi-label and binary settings.}
    \label{tab:feat}
\end{table}

\paragraph{Which feature is more predictive of addiction level? }We evaluate the feature importance on the test set to see their generalization ability through permutation feature importance~\cite{altmann2010permutation}.
After training the model, we randomly permute each feature from the test set and make predictions on the permuted data.
We then compute the difference between the macro $F_1$ score before and after permutation.
We repeat the process \num{1000} times for each feature.
\Cref{tab:feat} shows the drop in macro $F_1$ score after permutation.
In both settings, all the features contribute positively to the model's predictions.
Viewing time contributes the most to the predictions on unseen data, as demonstrated by the highest perturbed performance, followed by daily positive videos watched.
Under the multi-label setting, session size and gaps also contribute moderately to the prediction.
Overall, viewing time and positive videos watched are helpful features in the model's ability to generalize to unseen data.

\paragraph{Is viewing time sufficient for predicting addiction level?}

Viewing time being the most predictive feature organically leads to the question of whether viewing time on its own is sufficient to predict addiction level. 
To answer this, we train MLP models with only users' average viewing time as the independent feature and check its performance.


\begin{table}[h]
    \centering
    \small
    \begin{tabularx}{0.9\columnwidth}{{c} {r}{r}{r} {r}{r}{r}}
    \toprule
      \multirow{2}{*}{Class} &  \multicolumn{3}{{c}}{Multi} & \multicolumn{3}{{c}}{Binary} \\
         \cmidrule(lr){2-4} \cmidrule(lr){5-7}
        &  P    & R & $F_1$  &  P  & R & $F_1$ \\
    \midrule
     0 (35) & 0.42 & 0.43 & 0.42  & 0.71 & 0.82 & 0.76 \\ 
     1 (33) & 0.33 & 0.33 & 0.33  & 0.57 & 0.41 & 0.48 \\
     2 (39) & 0.53 & 0.51 & 0.52  &   \\
    \midrule 
    Macro Avg. & 0.43 & 0.42 & 0.43 & 0.64 & 0.62 & 0.62 \\
    \bottomrule
    \end{tabularx}
    \caption{Performance when using viewing time only.}
    \label{tab:ablation}
\end{table}

\Cref{tab:ablation} shows performance when the model is trained on viewing time only.
In both multi-label and binary settings, the overall performance drops (marginally) compared to using all the features in \Cref{tab:predict-model}.
Particularly in the multi-label setting, the model suffers from a significant drop in performance in identifying LLA and HLA users, as demonstrated by dropped recall. 
Further, to measure the unexplained variance in our classifiers' performances and to understand the potential difficulty of the underlying task, we calculate Brier score~\cite{brier1950verification} (lower the better).
In the multi-label setting, the Brier score is 0.65 for the classifier, with viewing time as the only feature.
It reduces to 0.60 with all the features, still implying a relatively high unexplained variance.

These observations further indicate that predicting one's addiction is rather complicated. 
Our evaluation may be seen as empirical evidence of social media addiction, not just about the amount of time a user spends on such platforms. 
Behavioral addiction on social media platforms may play out differently in individual cases, and more importantly, the user's social context (beyond that within the platform) may need to be understood for more accurate identification.

\paragraph{Main Takeaways of Section}
\begin{enumerate}
    \item Features derived from TikTok data can help identify HLA users from the rest.
    \item Usage patterns such as viewing time and positive video watched are helpful in the prediction task. 
   \item Predicting addiction level from basic usage patterns is rather difficult, and solely viewing time is insufficient. 
\end{enumerate}

\section{Ethical Considerations} 
\label{sec:ethics}

We conducted our study 
based on approval from the Ethical Review Board (ERB) of our university.
Before we collected a participant's TikTok data, we showed them a consent form and asked them to read the terms and conditions explained in the form and provide us with their explicit consent afterwards.
In order to protect the privacy of our participants, we establish stringent data privacy measures.
For instance, the donated TikTok dataset will not be shared with any third parties and is subject to complete deletion after 3 years of our project's completion.
When performing our analyses and reporting our results, we comply with standard ethical guidelines, such as reporting our results in aggregate~\cite{rivers2014ethical}.
Finally, as previously outlined in Section~\ref{sec:donation}, the video metadata collection is restricted to acquiring publicly available data on TikTok.
\section{Discussion \& Conclusion}
\label{sec:discussion_conclusion}

We conducted the first study on understanding end users' behavioral addiction on social media by surveying \num{1590} participants and stratifying them into three addiction groups.
For \num{107} participants, we obtained data donations and studied the extent to which different addiction groups exhibit platform usage and engagement patterns.
Finally, we designed a simple multi-layer perceptron model that, given basic social media usage features, achieves moderate performance in identifying highly likely addicted participants.
Next, we discuss our main findings and their implications, limitations, and potential future work.

\paragraph{Behavioral addiction in end users}
Our results show that 27\% of the surveyed participants are highly likely addicted to TikTok.
Furthermore, 39\% of participants from the age group $[18, 24]$ belong to this category, 
suggesting the prevalence of behavioral addiction particularly among young adults. It also suggests that the ``potential systemic risk''~\cite{EC2022DSARegulation} classification by DSA is justified as excessive social media use has been linked to 
negative effects on users' well-being~\cite{haand2020relationship}. 

\paragraph{Platform usage patterns}
Our results show that highly likely addicted users spend more time on the platform, and have shorter sessions gaps than less likely addicted users.
We argue that these findings were not a priori clear as harms, a prerequisite to being defined as addiction, could hypothetically be linked more to quality, not quantity of usage.

\paragraph{Predicting behavioral addiction}
By utilizing only basic social media usage data as our features, a multi-layer perceptron model achieves an $F_1$ score $\geq 0.55$ in identifying highly likely addicted users.
One main implication of this finding is that, although social media platforms have a moral obligation to identify at-risk users and take preventative measures to reduce the systemic risk, identifying at-risk users for addiction may be harder than anticipated based on digital traces. It may require additional attention to context beyond just social media usage.

\paragraph{Missing social context}

We speculate that a key missing feature is the \emph{social context}. Just as drinking alcohol with friends is, addiction-wise, less problematic than drinking alone, spending hours on TikTok to shoot, and share dance videos with friends might be less problematic than scrolling through TikTok to avoid meeting people in real life. Rich qualitative studies, potentially combined with group-based logging approaches, could help shed light on this.

\paragraph{Limitations}
First, our user sample is not representative with potential biases induced both by (i) the population of Prolific users, and (ii) self-selection into the survey. 
In the future, improved stratification methods could be used to more closely align the sampled population with a reference population.
Second, despite targeting a large audience for data donation, we obtained data from only 107 out of the \num{1590} surveyed participants.
In turn, this has led to partial demographic imbalances in our collected dataset.
We believe that the strong data quality checks we imposed, as well as the privacy-related concerns of the participants, contributed to this low turnout.
Third, our study only relies on video descriptions for the content analysis part. Utilizing different modalities, e.g., text, audio, video, etc., may improve the study further. 
Finally, the current study focuses on TikTok only, which prevents us from commenting about behavioral addiction across different social media platforms. 

\paragraph{Future Work}
One extension of our work could be to try and recruit a nationally representative set of survey (and data donation) participants. 
Replicating the same for several countries may also allow insights into cross-country differences and commonalities.
Another extension is to apply the same methodology to other social media platforms. This could help shed light on whether there are specifics in how TikTok operates that increase or decrease the risk of behavioral addiction.
Finally, it would be valuable to complement our quantitative methodology with qualitative, interview-based work. For example, sitting with a user and having them walk through some of their donated sessions could provide rich insight into what leads to problematic and excessive use of the app. 

\section{Acknowledgment}
\label{sec:ack}
Ingmar Weber and Cai Yang are supported by funding from the Alexander von Humboldt Foundation and its founder, the Federal Ministry of Education and Research (Bundesministerium für Bildung und Forschung).

\bibliography{aaai25}

\section{Paper Checklist}

\begin{enumerate}

\item For most authors...
\begin{enumerate}
    \item  Would answering this research question advance science without violating social contracts, such as violating privacy norms, perpetuating unfair profiling, exacerbating the socio-economic divide, or implying disrespect to societies or cultures?
    \answerYes{Yes, see section 1.}
  \item Do your main claims in the abstract and introduction accurately reflect the paper's contributions and scope?
    \answerYes{Yes, see sections 3, 5 and 6.}
   \item Do you clarify how the proposed methodological approach is appropriate for the claims made? 
    \answerYes{Yes, see sections 3 and 4.}
   \item Do you clarify what are possible artifacts in the data used, given population-specific distributions?
    \answerNo{No, the imbalance we observe is not part of the data collection process.}
  \item Did you describe the limitations of your work?
    \answerYes{Yes, see section 8.}
  \item Did you discuss any potential negative societal impacts of your work?
    \answerNA{NA}
      \item Did you discuss any potential misuse of your work?
    \answerNA{NA}
    \item Did you describe steps taken to prevent or mitigate potential negative outcomes of the research, such as data and model documentation, data anonymization, responsible release, access control, and the reproducibility of findings?
    \answerYes{Yes, see section 4.}
  \item Have you read the ethics review guidelines and ensured that your paper conforms to them?
    \answerYes{Yes}
\end{enumerate}

\item Additionally, if your study involves hypotheses testing...
\begin{enumerate}
  \item Did you clearly state the assumptions underlying all theoretical results?
    \answerNA{NA}
  \item Have you provided justifications for all theoretical results?
    \answerNA{NA}
  \item Did you discuss competing hypotheses or theories that might challenge or complement your theoretical results?
    \answerNA{NA}
  \item Have you considered alternative mechanisms or explanations that might account for the same outcomes observed in your study?
    \answerNA{NA}
  \item Did you address potential biases or limitations in your theoretical framework?
    \answerNA{NA}
  \item Have you related your theoretical results to the existing literature in social science?
    \answerNA{NA}
  \item Did you discuss the implications of your theoretical results for policy, practice, or further research in the social science domain?
    \answerNA{NA}
\end{enumerate}

\item Additionally, if you are including theoretical proofs...
\begin{enumerate}
  \item Did you state the full set of assumptions of all theoretical results?
    \answerNA{NA}
	\item Did you include complete proofs of all theoretical results?
    \answerNA{NA}
\end{enumerate}

\item Additionally, if you ran machine learning experiments...
\begin{enumerate}
  \item Did you include the code, data, and instructions needed to reproduce the main experimental results (either in the supplemental material or as a URL)?
    \answerNo{No, we are using open-source classifiers on scikit-learn. Our dataset contains data about individual users, and we choose not to release them for privacy concerns.}
  \item Did you specify all the training details (e.g., data splits, hyperparameters, how they were chosen)?
    \answerYes{Yes, see section 6.}
     \item Did you report error bars (e.g., with respect to the random seed after running experiments multiple times)?
    \answerYes{Yes, see section 6.}
	\item Did you include the total amount of compute and the type of resources used (e.g., type of GPUs, internal cluster, or cloud provider)?
    \answerNo{No, we use very simple machine learning models which are not limited to computational resources.}
     \item Do you justify how the proposed evaluation is sufficient and appropriate to the claims made? 
    \answerYes{Yes, see section 6.}
     \item Do you discuss what is ``the cost`` of misclassification and fault (in)tolerance?
    \answerYes{Yes, see section 6.}
  
\end{enumerate}

\item Additionally, if you are using existing assets (e.g., code, data, models) or curating/releasing new assets, \textbf{without compromising anonymity}...
\begin{enumerate}
  \item If your work uses existing assets, did you cite the creators?
    \answerNA{NA}
  \item Did you mention the license of the assets?
    \answerNA{NA}
  \item Did you include any new assets in the supplemental material or as a URL?
    \answerNA{NA}
  \item Did you discuss whether and how consent was obtained from people whose data you're using/curating?
    \answerYes{Yes, see section 7.}
  \item Did you discuss whether the data you are using/curating contains personally identifiable information or offensive content?
    \answerYes{Yes, see section 4.}
\item If you are curating or releasing new datasets, did you discuss how you intend to make your datasets FAIR (see \citet{fair})?
\answerNA{NA}
\item If you are curating or releasing new datasets, did you create a Datasheet for the Dataset (see \citet{gebru2021datasheets})? 
\answerNA{NA}
\end{enumerate}

\item Additionally, if you used crowdsourcing or conducted research with human subjects, \textbf{without compromising anonymity}...
\begin{enumerate}
  \item Did you include the full text of instructions given to participants and screenshots?
    \answerYes{Yes, see section 4 and appendix A.2.}
  \item Did you describe any potential participant risks, with mentions of Institutional Review Board (IRB) approvals?
    \answerYes{Yes, see section 7.}
  \item Did you include the estimated hourly wage paid to participants and the total amount spent on participant compensation?
    \answerYes{Yes, see sections 3 and 4.}
   \item Did you discuss how data is stored, shared, and deidentified?
   \answerYes{Yes, see sections 4 and 7.}
\end{enumerate}

\end{enumerate}

\appendix
\section{Appendix}

\subsection{Example survey questions}
\label{app:question}

\noindent
\Cref{tab:survey-question} displays five examples of questions from each section of the survey.

\begin{table}[h]
    \centering
    \small
    \begin{tabular}{p{0.33\columnwidth} p{0.57\columnwidth}}
    \toprule
     \textbf{Demographics} & \textbf{TikTok usage habits} \\
    \midrule
     What is your gender? & I consider myself addicted to TikTok \\ 
    \midrule
     What is your employment status? & How many hours per day do you spend on TikTok? \\
    \midrule
    In which country are you located now? & How immediately do you check TikTok once you receive any notification? \\
    \midrule
    When do you usually go to bed? & I become restless or troubled if I am unable to use TikTok \\
    \midrule
    When do you usually wake up? & How often do you use TikTok per day? \\
    \bottomrule
    \end{tabular}
    \caption{Example demographics and TikTok usage habits questions from our survey.}
    \label{tab:survey-question}
\end{table}

\subsection{TikTok data request instructions}
\label{app:instruction}
\begin{figure*}[ht]
  \centering
  \includegraphics[width=2\columnwidth]{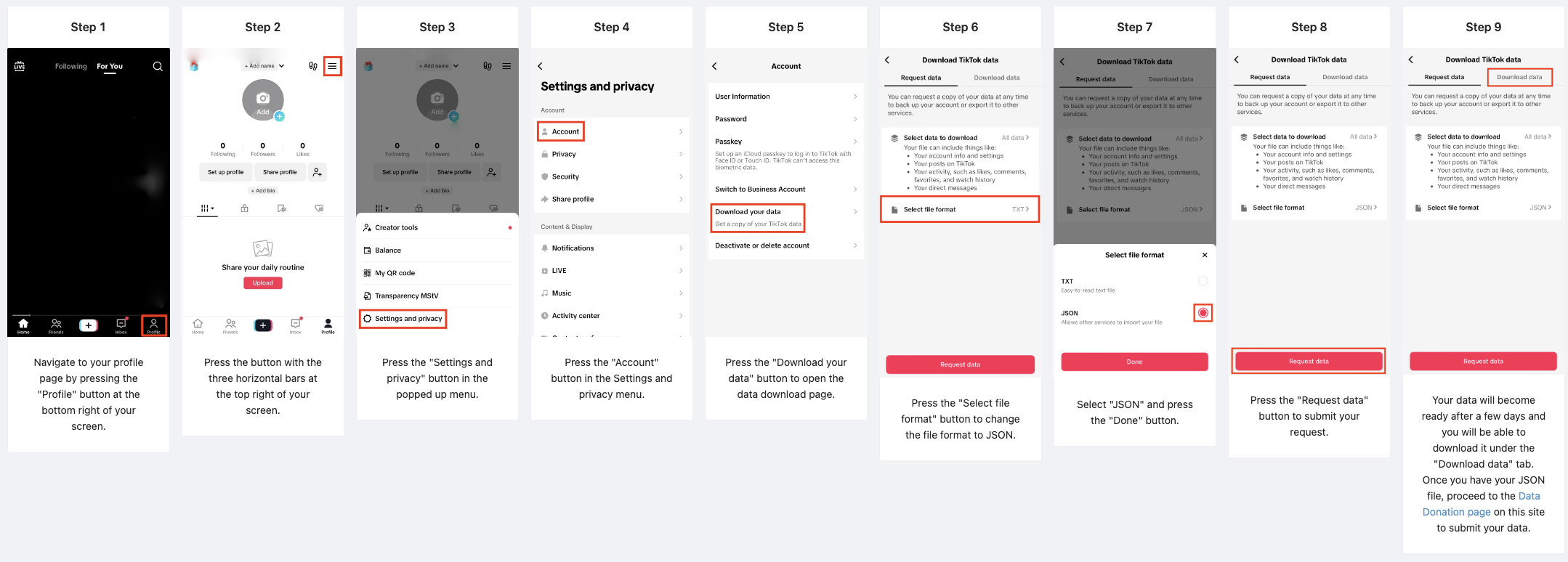}
  \caption{Instructions provided to different users on our data donation platform.}
  \label{fig:instruction}
\end{figure*}

\Cref{fig:instruction} displays the instructions to request data copies on TikTok.

\subsection{Gender, age and mobile OS breakdown by addiction group}
\label{app:demo-survey}

\noindent
\Cref{tab:demo-from-survey} reports the gender, age, and mobile operation system (OS) of each addiction group.
HLA and MLA have marginally female participants (53.4\% and 55.7\%), while LLA consists of more male participants (57.8\%).
The age group is shifting from young adults to middle-aged people with decreasing addiction levels.
More than half of the participants in each group are using iOS on their mobile.

\begin{table}[h]
    \centering
    \small
    \begin{tabular}{crcccc}
    \toprule
           &  & LLA & MLA & HLA & Total \\
    \midrule
    \multirow{3}{*}{Gender} & Female & 275 & 266 & 233 & 774 \\
    & Male & 378 & 212 & 195 & 785 \\
    & Non-binary & 11 & 8 & 8 & 27\\
    & Prefer not to say & 3 & 1 & 0 & 4 \\
    \midrule
    \multirow{4}{*}{Age} & 18-24 & 119 & 108 & 144 & 371 \\
    & 25-34  & 236 & 200 & 175 & 611 \\
    & 35-44 & 155 & 92 & 52 & 299 \\
    & 45-64  & 142 & 80 & 60 & 282 \\
    & 65+ & 14 & 6 & 5 & 25 \\
    & Prefer not to say & 1 & 1 & 0 & 2\\
    \midrule
    \multirow{2}{*}{Mobile OS} & iOS & 380 & 319 & 315 & 1014 \\
    & Android & 284 & 165 & 119 & 568 \\
    & Prefer not to say & 3 & 3 & 2 & 8\\
    \bottomrule
    \end{tabular}
    \caption{Gender, age, and mobile operating system distribution within the three addiction groups for all the participants from the survey. Total number is shown in the rightmost column. }
    \label{tab:demo-from-survey}
\end{table}

\subsection{Bergen Facebook Addiction Scale under the context of TikTok}
\label{app:BFAS}

\noindent
\Cref{tab:bfas} displays the set of six questions from Bergen Facebook Addiction Scale~\cite{andreassen2015online} adapted for TikTok.

\begin{table}[h]
    \centering
    \begin{tabularx}{0.99\columnwidth}{p{3cm}X X}
    \toprule
    Salience & I spent a lot of time thinking about TikTok or planned use of TikTok \\
    \midrule
    Tolerance & I feel an urge to use TikTok more and more \\ 
    \midrule
    Mood Modification & I have used TikTok in order to forget about personal problems \\
    \midrule
    Relapse & I have tried to cut down on the use of TikTok without success \\
    \midrule
    Withdrawal & I become restless or troubled if I am unable to use TikTok \\
    \midrule
    Conflict & I have used TikTok so much that it has had a negative impact on my job/studies \\
    \bottomrule
    \end{tabularx}
    \caption{Six items from BFAS adapted into the context of TikTok.}
    \label{tab:bfas}
\end{table}

\subsection{Implausible answers to certain questions}
\label{app:implausible}

\noindent
\Cref{tab:implausbile-answers} shows the set of curated implausible question and answer pairs.

\begin{table}[ht]
    \centering
    \begin{tabularx}{0.99\columnwidth}{p{3.5cm}X X}
    \toprule
    What is your age? What is your marital status? & (18-24 years old, Widowed); (18-24 years old, Divorced); (18-24 years old, Separated) \\
    \midrule
    How often do you use TikTok per day? How many hours per day do you spend on TikTok & ($<$ 5 times, $>$ 8 hours); ($<$ 5 times, 4-8 hours); ($<$ 50 times, $<$ 1 hour) \\
    \midrule
    How many hours per day do you spend on TikTok? How frequently do you check TikTok within the first 30 minutes of waking up in a day? & ($>$ 8 hours, Never); ($>$ 8 hours, Rarely); ($<$ 1 hour, Always) \\
    \midrule
    How often do you use TikTok per day? How frequently do you check TikTok within the first 30 minutes of waking up in a day? & ($>$ 50 times, Never); ($>$ 50 times, Rarely) \\
    \bottomrule
    \end{tabularx}
    \caption{Question pairs and implausible answers.}
    \label{tab:implausbile-answers}
\end{table}

\subsection{Breakdown of the answers to the explicit addiction question by addiction group}
\label{app:explicit-question}

\begin{figure}[ht]
  \centering
  \includegraphics[width=0.99\columnwidth]{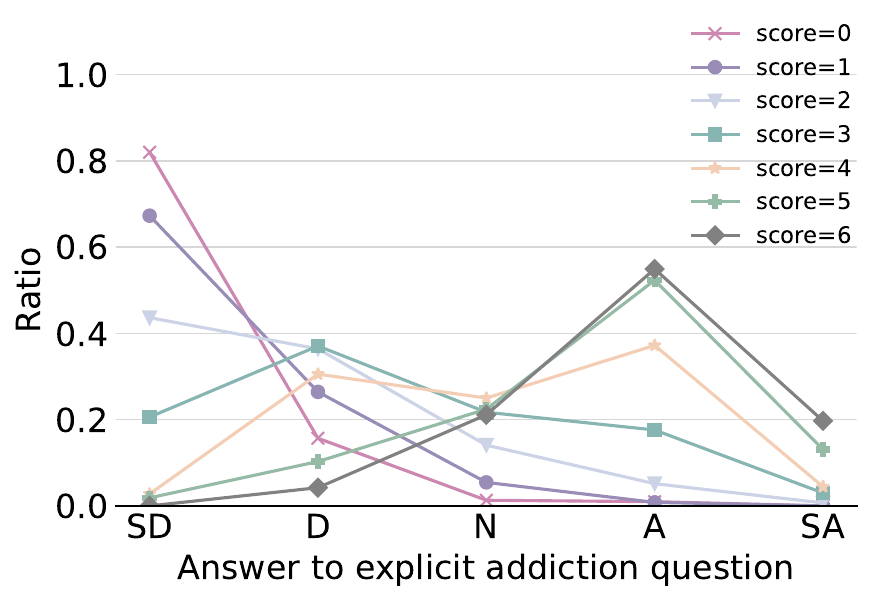}
  \caption{Participants' answers for the explicit addiction question across different addiction scores. The x-axis labels stand for strongly disagree, disagree, neutral, agree, and strongly agree, respectively. Participants with higher addiction scores are more likely to claim self-addiction.}
  \label{fig:psy-vs-explicit}
\end{figure}

\noindent
\Cref{fig:psy-vs-explicit} shows how participants responded to the explicit addiction question depending on their addiction scores for all possible 7 addiction scores.
We find that less than 1\% of participants with addiction scores of 0 and 1 respond positively to the explicit addiction question.
This percentage is 13\% for addiction scores of 2 and 3, and it goes up to 55\% for participants with addiction scores of 4 or above.
We observe an increasing trend to respond positively to the explicit addiction when the addiction score goes up.
Meanwhile, we also observe that 71\% of participants who self-report as being addicted have an addiction score of 4 or above.
In a nutshell, we find that participants with higher BFAS-based addiction scores are more likely to self-report being addicted and vice versa, indicating consistency between the results of the psychological method and the participants' perception of their behavioral addiction level.

\subsection{Data Completeness Check}
\label{app:data-complete}

\noindent
Since our analyses will rely on the data provided by TikTok to the corresponding users, it is imperative to understand the completeness and correctness of such a data sample. To that end, we manually check the completeness of data (videos) returned by TikTok for subsets of our own data.
To do so, three authors watched videos on TikTok for about 10 minutes. 
They also manually interacted with videos, e.g., by liking or favoring them.
They screen-recorded the videos watched, requested their data and then compared the returned content.
We find that the data returned by TikTok captures all the video interactions exactly, which leads us to believe that the donated TikTok data is complete regarding videos watched.

\subsection{Comparing participants' responses with the digital data }
\label{app:data-survey-compare}

\noindent
\Cref{subsec:related-survey-log} showed that self-reports can be overestimates or underestimates of the actual use.
We compare the reported amount of time and frequency of usage with participants' donated data.
For the time spent on TikTok, we find that about 60\% of users’ actual time spent (as per log data) is within their self-reported usage intervals (up to a tolerance of $\pm$ 25\%).
Of the remaining 40\% of users, almost everyone is over-reporting their time spent on TikTok, i.e., self-reported time $>$ logged time.

For the frequency of TikTok, we find that 74\% of users’ actual frequency, as shown by their data, is aligned with their self-reports, using the same methodology.
Around 67\% of the other users are under-reporting their actual TikTok usage frequency.

In conclusion, most users can roughly estimate their usage, but there is a tendency to (i) overestimate the time spent on the platform and (ii) underestimate the usage frequency.

We also checked Pearson's $r$ between their digital data and self-reports and found a low to moderate correlation in both cases.
All of these findings are in line with existing works~\cite{parry2021systematic,ernala2020well,goetzen2023likes}.

The potential reasons for such misalignment could be attributed to factors such as demographics, engagement, or platform algorithms, the investigation of which is out of the scope of this paper.

However, we want to highlight that we are not relying on users’ answers to the objective questions and, e.g., the classifier uses the logged behavioral data. At the same time, we have to rely on the participants’ subjective feelings toward their usage of TikTok for the addiction-related questions as this is the standard way to measure people's addiction. Unfortunately, notions such as “harms” are not evident in the log data. In summary, we believe proceeding with users’ self-reports on the addiction-related questions is largely unavoidable.

\subsection{  Video viewing duration }
\label{app:halfaker-viewing}

\begin{figure}[t]
  \centering
  \includegraphics[width=0.99\columnwidth]{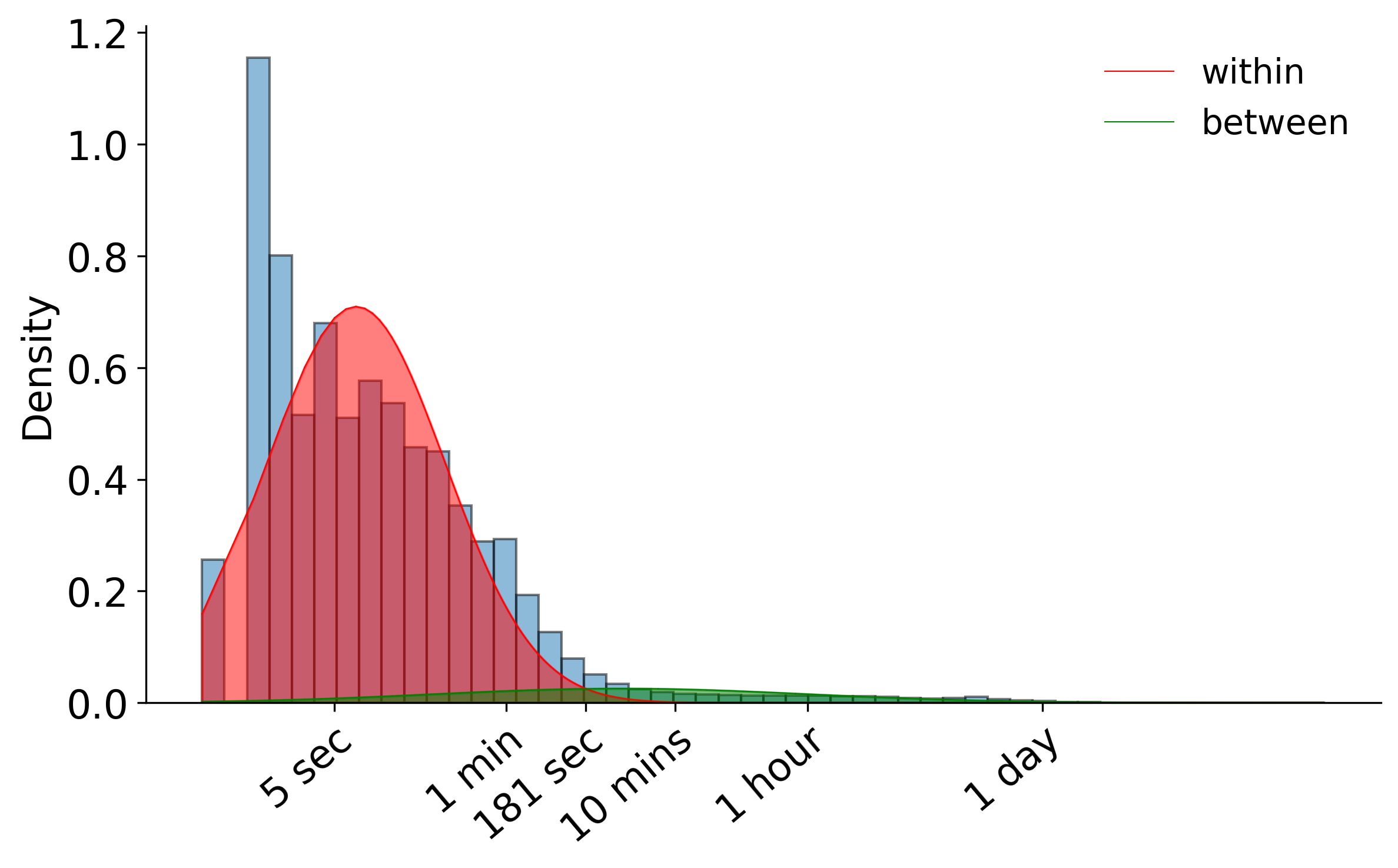}
  \caption{ Histogram of viewing duration in logarithm. Curves represent bimodal clusters output from Gaussian Mixture Models.}
  \label{fig:halfaker}
\end{figure}

\noindent
\Cref{fig:halfaker} shows the logarithm of inferred viewing duration and fitted Gaussian Mixture Models.
The two mixture models intersect at 181 seconds.
An inferred viewing duration longer than 181 seconds indicates a potential session break.

\subsection{ Effects of BFAS questions on observations }
\label{app:question-check}

Among BFAS questions, Conflict, Salience and Relapse may be considered to be ex- or implicitly related to usage time.
If, based on the questions to these answers, users are then labeled as HLA, then it is trivial that higher addiction levels relate to higher usage intensity. 
Here, we outline why we do not think that this concern is warranted, discussing each of the three dimensions, one after the other.

\noindent
\textbf{(i) Conflict}: this dimension might indirectly relate to app usage as the survey question explicitly mentions ``I have used TikTok so much that ...''.
However, empirically, we observe that (1) this link is not very pronounced, and, importantly, (2) the contribution to labeling users as HLA is minimal.
We compute the percentage of times HLA users responded with \textit{Very Often}, \textit{Often} and \textit{Sometimes} to the BFAS questions.  
The sorted percentages for Mood Modification, Tolerance, Salience, Relapse, Conflict and Withdrawal are 97\%, 92\%, 85\%, 69\%, 41\% and 33\%, respectively (As a reference, the numbers for all the surveyed participants are 68\%. 49\%, 38\%, 27\%, 21\%, 12\%, and for all the 107 participants, whose data donations were studied, are 77\%, 65\%, 50\%, 28\%, 19\%, 15\%). 
Mood Modification, Tolerance, Salience and Relapse are the top four options with the most positive responses. 
This finding holds for both the entire survey population (N=\num{1590}) and the 107 users who donated their data.
It means that these four dimensions contribute far more to the classification of users as HLA than the other two dimensions (Conflict and Withdrawal).

To further demonstrate that the effect of the answers to the Conflict question on the user classification is small, we have manually set every user's response to Conflict to be negative (\textit{Very Rarely}). 
Under this new setting, the addiction classification changes only slightly from the original 39 HLA, 33 MLA and 35 LLA, to 35 HLA, 36 MLA and 36 LLA.
The classification of most users from each group remains intact: the overlap between the new and original group is 35 for HLA (90\%), 32 for MLA (97\%), and 35 for LLA (100\%), respectively. 
However, even if we were to consider these changes to be big, then the overall results reported would still not change. 
Our previous observations in Section 5 all stand under this new setting. 
We report the new measurements below.
For simplicity, we only report metrics that were previously shown to be statistically different across groups.

For viewing time (in hours), HLA: 1.48 $\pm$ 0.14, MLA: 1.02 $\pm$ 0.09, LLA: 0.88 $\pm$ 0.11. 
HLA users have a higher average viewing time than the other two groups (HLA and MLA: $p < 0.01$, Cohen's $d = 0.87$; HLA and LLA: $p < 0.01$, Cohen's $d = 0.82$).
For session size, HLA: 29.48 $\pm$ 1.99, MLA: 23.59 $\pm$ 2.04, LLA: 32.16 $\pm$ 4.06. 
MLA users have watched fewer videos each time than the other two groups (MLA and HLA: $p < 0.05$, Cohen's $d = 0.22$; MLA and LLA: $p < 0.05$, Cohen's $d = 0.14$).
For session gaps (in hours), HLA: 3.99 $\pm$ 0.91, MLA: 4.67 $\pm$ 0.59, LLA: 8.89 $\pm$ 1.57. 
LLA users have returned less frequently to TikTok (LLA and HLA: $p < 0.01$, Cohen's $d = 0.65$; LLA and MLA: $p < 0.01$, Cohen's $d = 0.70$).
For usage during the daytime, HLA: 0.43 $\pm$ 0.16, MLA: 0.51 $\pm$ 0.15, LLA: 0.51 $\pm$ 0.18. 
HLA users have used TikTok more at night (HLA and LLA: $p < 0.05$, Cohen's $d = 0.48$; HLA and MLA: $p < 0.05$, Cohen's $d = 0.51$).

Overall, we observe very similar results to those reported in the main draft. 
This observation demonstrates that Conflict, even if assumed to be related to app usage time, only slightly changes the set of users classified as HLA, and does not impact our results significantly.

\noindent

\textbf{(ii) Salience}: we argue that there is a distinction between a user ``spending lots of time \textit{thinking} about TikTok'' and ``spending lots of time \textit{using} TikTok''. 
A positive response to the former indicates emotional attachment to the platform, but it does not necessarily lead to excessive usage.
In our opinion, observing a link between the emotional attachment and the actual usage time is non-obvious. 
Empirically, 11 HLA users answer positively to the Salience question (= high emotional attachment) but spend relatively little time on TikTok ($<$ 0.84 hours per day, lower than the mean for LLA). 
3 HLA users spend a relatively high amount of time on TikTok ($>$ 1.45 hours per day, higher than the mean for HLA) but respond negatively to this question. 
6 LLA users answer negatively to this question but spend a relatively high amount of time on TikTok ($>$ 1.45 hours per day).

\noindent

\textbf{(iii) Relapse}: similarly, we argue that failing to cut down the time on TikTok does not indicate high usage time either. 
Instead, it reflects the behavioral dependence that the user has on TikTok.
Empirically, similar to Salience, there are 7 HLA users who answer positively to the Relapse question but spend a relatively lower amount of time on TikTok than the average of LLA ($<$ 0.84 hours per day). 
There are 6 HLA users who claim no failure to cut down their usage time but still spend a relatively high amount of time ($>$ 1.45 hours per day, higher than the mean for HLA) on TikTok. 
There are also 6 LLA users claiming no failure but spending a relatively high amount of time ($>$ 1.45 hours per day).

In short, we do not see a strong theoretical and obvious link between Relapse, Salience and usage time. 
Empirically, there are users who, indeed, demonstrate one without the other. 
For completeness, we also repeat the experiments on Salience and Relapse.
For simplicity, we use $>$ and $<$ to represent the comparison between addiction groups.

Results on Salience (31 HLA, 37 MLA, 39 LLA after change; the overlap is 31 for HLA (79\%), 29 (88\%) for MLA, 35 (100\%) for LLA):
For viewing time, HLA: 1.53 $\pm$ 0.15, MLA: 1.07 $\pm$ 0.10, LLA: 0.85 $\pm$ 0.10. 
HLA $>$ MLA ($p < 0.01$, Cohen's $d = 0.97$) and HLA $>$ LLA ($p < 0.01$, Cohen's $d = 0.95$).
For session size, HLA: 28.67 $\pm$ 1.89, MLA: 25.04 $\pm$ 2.25, LLA: 31.38 $\pm$ 3.80. 
We find a p-value between $0.05$ and $0.1$ when comparing MLA and LLA users.
Although no statistical significance exists, it still suggests a potential trend that MLA users watch fewer videos than LLA users.
For session gaps, HLA: 3.79 $\pm$ 0.97, MLA: 4.52 $\pm$ 0.62, LLA: 8.79 $\pm$ 1.45. 
LLA $<$ HLA ($p < 0.01$, Cohen's $d = 0.66$) and LLA $<$ MLA ($p < 0.01$, Cohen's $d = 0.73$).
For usage during the daytime, HLA: 0.43 $\pm$ 0.17, MLA: 0.50 $\pm$ 0.15, LLA: 0.51 $\pm$ 0.17. 
HLA $<$ LLA ($p < 0.05$, Cohen's $d = 0.45$) and HLA $<$ MLA ($p < 0.05$, Cohen's $d = 0.49$).

Results after changing responses to Relapse (32 HLA, 36 MLA, 39 LLA after change; the overlap is 32 for HLA (82\%), 32 (97\%) for MLA, 35 (100\%) for LLA):
For viewing time, HLA: 1.56 $\pm$ 0.15, MLA: 1.01 $\pm$ 0.09, LLA: 0.86 $\pm$ 0.10. 
HLA $>$ MLA ($p < 0.01$, Cohen's $d = 1.03$) and HLA $>$ LLA ($p < 0.01$, Cohen's $d = 0.98$).
For session size, HLA: 29.37 $\pm$ 2.20, MLA: 23.73 $\pm$ 1.86, LLA: 32.60 $\pm$ 4.05. 
MLA $<$ HLA ($p < 0.05$, Cohen's $d = 0.27$) and MLA $<$ LLA ($p < 0.05$, Cohen's $d = 0.17$). 
For session gaps, HLA: 3.68 $\pm$ 0.94, MLA: 4.77 $\pm$ 0.59, LLA: 9.00 $\pm$ 1.56. 
LLA $<$ HLA ($p < 0.01$, Cohen's $d = 0.70$) and LLA $<$ MLA ($p < 0.01$, Cohen's $d = 0.77$).
For usage during the daytime, HLA: 0.43 $\pm$ 0.17, MLA: 0.50 $\pm$ 0.14, LLA: 0.51 $\pm$ 0.18. 
HLA $<$ LLA ($p < 0.05$, Cohen's $d = 0.47$) and HLA $<$ MLA ($p < 0.05$, Cohen's $d = 0.52$).

In conclusion, we believe our experiments outlined above have demonstrated that the questions used to classify addiction are not related to app usage time, or if they are indirectly related, the effect is minimal.

\subsection{Examples of video descriptions with positive sentiment}
\label{app:positive-example}

\noindent
We list five example video descriptions identified to have positive sentiment:
\begin{enumerate}
    \item Holu Steakhouse was amazing! If you want a hidden gem for a date night check it out! \#fyp \#finedining \#makeitcinematic \#blackwomeninluxury
    \item I can’t recommend this product enough it is literally life changing \#acne \#bodyacne \#acnesolution
    \item We're just groovin' this Saturday! Hope you all have an amazing weekend! \#FitFun \#bungeefitness \#bungeeworkout \#BungeeONE \#fitness \#FlyHigh
    \item Happy Holidays and have a times of family \#julesleblanc \#jaydenbartels \#sidehustle
    \item Big thanks to Luna The Panteras Parents for letting us film these awesome clips of the most beautiful panther! \#panther \#panthers \#petsoftiktok \#exoticpets \#luna \#rottweiler \#rottweilersoftiktok
\end{enumerate}

\subsection{Addiction prediction using different input sources}
\label{app:predict-input}

\noindent
We have experimented with different data sources in the addiction prediction task using MLP models, based on the same procedure for hyperparameter turning and feature selection.
The motivation behind this is to see the changes in performance when having access to various data sources.

\begin{table}[h]
    \centering
    \scriptsize
    \begin{tabular}{cc rrr rrr}
    \toprule
      \multirow{2}{*}{Input} & \multirow{2}{*}{Class} &  \multicolumn{3}{c}{Multi} & \multicolumn{3}{c}{Binary } \\
         \cmidrule(lr){3-5} \cmidrule(lr){6-8}
       & &  P    & R & $F_1$  &  P  & R & $F_1$ \\
        \midrule
    \multirow{3}{*}{Survey} & 0 & 0.62 & 0.69 & 0.65  & 0.74 & 0.82 & 0.78 \\ 
    & 1 & 0.27 & 0.21 & 0.24  & 0.61 & 0.49 & 0.54 \\
    & 2 & 0.55 & 0.59 & 0.57  &   \\    
    \cmidrule(lr){2-8}
    & Macro Avg. & 0.48 & 0.50 & 0.48 & 0.67 & 0.66 & 0.66 \\
    
    \midrule
    
    Survey & 0 & 0.57 & 0.57 & 0.57   &  0.81 & 0.85 & 0.83 \\ 
     + & 1 & 0.31 & 0.24 & 0.27 & 0.71 & 0.64 & 0.68  \\
    Data & 2 & 0.54 & 0.64 & 0.59 &   \\

    \cmidrule(lr){2-8}
    & Macro Avg. & 0.47 & 0.48 & 0.48 & 0.76 & 0.75 & 0.75 \\
    
    \bottomrule
    \end{tabular}
    \caption{Addiction prediction performance of MLP models using different input sources. In the multi-label setting, class 0/1/2 represents LLA/MLA/HLA respectively. In the binary setting, class 0 represents LLA+MLA while class 1 represents HLA.}
    \label{tab:predict-model-appendix}
\end{table}


As seen in \Cref{fig:usage-question-vs-addict}, users from different addiction groups tend to answer differently about their TikTok usage questions. 
We use their answers to these questions as input to the models. 
We focus on their responses to three questions: (1) How frequently do you check TikTok within the first 30 minutes of waking up in a day? (2) How often do you use TikTok per day? And (3) How many hours per day do you spend on TikTok?
We also combine the same set of features in \Cref{tab:feat} with the aforementioned survey answers as the input to models.

\Cref{tab:predict-model-appendix} reports the performance of MLP models with survey features only (Survey) and using combined features (Survey + Data).
The performance differs marginally between the two input sources.
Under binary settings, performance using all three sources is close.
Under multi-label settings, the overall performance is slightly better when using digital data only.


\subsection{Addiction prediction}
\label{app:predict}

\Cref{tab:predict-survey,tab:predict-survey-data,tab:predict-data} shows the cross-validation performance on all the models we have experimented with in this paper using different input features.

\begin{table}[h]
    \centering
    \begin{tabular}{rrrrr}
    \toprule
        & \multicolumn{2}{c}{Multi} & \multicolumn{2}{c}{Binary} \\
        
        \cmidrule(lr){2-3}  \cmidrule(lr){4-5} 
        & Validation      & Test    & Validation      & Test \\
        \midrule
    DT  & \textbf{0.55$\pm$0.16} & \underline{0.48}  & \textbf{0.76$\pm$0.12} & 0.65      \\
    LR  & 0.50$\pm$0.15 & \textbf{0.53}  & \underline{0.72$\pm$0.11} & 0.64  \\
    SVM & 0.47$\pm$0.14 & 0.42  & 0.70$\pm$0.11 & 0.64    \\
    KNN & 0.51$\pm$0.12 & \textbf{0.53}  & 0.70$\pm$0.13 & \textbf{0.71}      \\
    MLP & \underline{0.54$\pm$0.14} & \underline{0.48}  & \underline{0.72$\pm$0.11} & \underline{0.66}   \\
    \bottomrule
    \end{tabular}
    \caption{Macro $F_1$. using only survey features. \textbf{Bold}: best performance; \underline{underline}: second best performance.}
    \label{tab:predict-survey}
\end{table}

\begin{table}[h]
    \centering
    \begin{tabular}{rrrrr}
    \toprule
        & \multicolumn{2}{c}{Multi} & \multicolumn{2}{c}{Binary} \\
        
        \cmidrule(lr){2-3}  \cmidrule(lr){4-5} 
        & Validation    & Test    & Validation    & Test \\
        \midrule
    DT  & \textbf{0.56$\pm$0.14} & 0.41  & 0.72$\pm$0.14 & 0.63   \\
    LR  & 0.52$\pm$0.16 & \textbf{0.53}  & \textbf{0.75$\pm$0.15} & \underline{0.72}  \\
    SVM & 0.49$\pm$0.16 & 0.46  & \underline{0.74$\pm$0.15} & 0.68   \\
    KNN & 0.53$\pm$0.15 & \textbf{0.53}  & 0.69$\pm$0.15 & 0.69   \\
    MLP & \underline{0.54$\pm$0.16} & \underline{0.48}  & \textbf{0.75$\pm$0.14} & \textbf{0.75}  \\
    \bottomrule
    \end{tabular}
    \caption{Macro $F_1$. using survey and data features. \textbf{Bold}: best performance; \underline{underline}: second best performance.}
    \label{tab:predict-survey-data}
\end{table}

\begin{table}[h]
    \centering
    \begin{tabular}{rrrrr}
    \toprule
        & \multicolumn{2}{c}{Multi} & \multicolumn{2}{c}{Binary} \\
        
        \cmidrule(lr){2-3}  \cmidrule(lr){4-5} 
        & Validation    & Test    & Validation    & Test \\
        \midrule
    DT  & 0.49$\pm$0.18 & 0.47 & \underline{0.65$\pm$0.15} & 0.57  \\
    LR  & \textbf{0.53$\pm$0.18} & \textbf{0.55} & \textbf{0.67$\pm$0.22} & \underline{0.65}  \\
    SVM & \textbf{0.53$\pm$0.18} & \underline{0.53} & 0.63$\pm$0.23 & 0.64 \\
    KNN & 0.47$\pm$0.17 & 0.46 & 0.60$\pm$0.19 & 0.54  \\
    MLP & \underline{0.51$\pm$0.18} & \underline{0.53} & \textbf{0.67$\pm$0.21} & \textbf{0.67}  \\
    \bottomrule
    \end{tabular}
    \caption{Macro $F_1$. using data features. \textbf{Bold}: best performance; \underline{underline}: second best performance.}
    \label{tab:predict-data}
\end{table}



\end{document}